\documentclass[twocolumn,aps,prl,showpacs,amsmath,superscriptaddress]{revtex4-1}
\usepackage{amssymb}
\usepackage{mathrsfs}
\usepackage{graphicx}
\usepackage{float}
\usepackage[caption=false]{subfig}

\usepackage{epstopdf}

\usepackage[normalem]{ulem}
\usepackage{color}
\usepackage{bm}

\begin{document}
\title{Realistic Floquet semimetal with exotic topological linkages between arbitrarily many nodal loops}

\author{Linhu Li}
\affiliation{Department of Physics, National University of Singapore, Singapore 117551, Republic of Singapore}
\author{Ching Hua Lee}
\email{calvin-lee@ihpc.a-star.edu.sg}
\affiliation{Department of Physics, National University of Singapore, Singapore 117551, Republic of Singapore}
\affiliation{Institute of High Performance Computing, Singapore 138632, Republic of Singapore}
\author{Jiangbin Gong}
\email{phygj@nus.edu.sg}
\affiliation{Department of Physics, National University of Singapore, Singapore 117551, Republic of Singapore}

\begin{abstract}
 Valence and conduction bands in nodal loop semimetals (NLSMs) touch along closed loops in momentum space. If such loops can proliferate and link intricately, NLSMs become exotic topological phases, 
 which require non-local hopping and are therefore unrealistic in conventional quantum materials or cold atom systems alike.  In this work, we show how this hurdle can be surmounted through an experimentally feasible periodic driving scheme. In particular, by tuning the period of a two-step periodic driving or certain experimentally accessible parameters, we can  generate arbitrarily many nodal loops that are linked with various levels of complexity. Furthermore, we propose to use both a Berry-phase related winding number and the Alexander polynomial topological invariant to characterize the fascinating linkages among the nodal loops. This work thus presents a class of exotic Floquet topological phases that has hitherto not been proposed in any realistic setup. {Possible experimental confirmation of such exotic topological phases is also discussed.}
\end{abstract}
\maketitle



Nodal loop semimetals (NLSMs) are 3D systems with valence and conduction bands touching along closed loops in momentum space
 \cite{Burkov_2011,Fan1,Fan2,Weng_2015,Zhang_2016}.
{They feature flat two-dimensional (2D) ``drumhead" surface states protected by the bulk topology, and are hence characterized by anomalously large surface density of states} that give rise to interesting transport properties \cite{transport_NL1,transport_NL2,transport_NL3,transport_NL4,lee2015negative}. {NLSMs can be used to generate many gapped and gapless topological phases upon breaking certain symmetries} \cite{Kim_2015,Yu_2015,driven_NL1,driven_NL2,driven_NL3,driven_NL4,Yan_2016,Yan_2017,Li2017,Li2017_3}. {Nodal loops with superficially similar structures may exhibit markedly different topological characteristics}, e.g.  $Z_2$ monopole loops \cite{Fang_2015}, vortex rings \cite{Lim_2017}, and $2\pi$-flux loops \cite{Li2017_2}.

{Compared} to the well-studied nodal point semimetals (NPSMs) with zero-dimensional (0D) band touching points \cite{Wan_2011,Young_2012}, {NLSMs possess far richer topological features. Indeed, nodal loops as closed 1-dimensional (1D) manifolds in 3D momentum space can be knotted or linked in an infinite number of qualitatively distinct ways \cite{Zhong_2017,Ezawa_Hopf,Yan_Hopf,Chen_Hopf,XGWan}.
The classification of nodal knots or links involves not just a simple group like $\mathbb{Z}_2$ or $\mathbb{Z}$ as in NPSMs, Chern and spin Hall insulators etc., but is so inherently complicated that no single topological invariant can unambiguously distinguish all the possible inequivalent manifestations {with the same number of links}. For example, three nodal loops can be non-trivially linked even when none of them are linked pair-wise, a situation that can only be identified with the Milnor invariant or other link polynomials. Physically, intricately knotted or linked nodal loops can yield unconventional transport behavior, such as negative differential resistance at the nodal loop boundaries~\cite{lee2015negative} and a possible topological shift in the coefficient of thermal
magnetoelectric effect \cite{Sun17}.  The simplest {physical realizations} of a pair of Hopf-link has been proposed in Co$_2$MnGa in a 4-band model~\cite{4band_Hopf}, and in superconducting circuits where the nodal links are emulated in an effective 3D parameter space rather than momentum space \cite{SC_circuits}. As a more subtle case, NLSMs with nodal chains have also been experimentally realized in metallic-mesh photonic crystals~\cite{yan2018experimental}.}  However,  NLSM with multiply linked nodal loops or knots have remained elusive.

Recent studies of Floquet topological matter \cite{Floquet1,Floquet2,Floquet3,Floquet4,Floquet5,Floquet6,Floquet7,Floquet8,Floquet9,Floquet10,Floquet_MJ1,Floquet_MJ2,Floquet_MJ4,Floquet_SM1,Floquet_SM2,exp_PL1,exp_PL2,exp_PL3,exp_PL4,exp_OL1,exp_OL2} have discovered rich topological phases with unusually large topological invariants \cite{Tong_2013,Derek_2014,Tianshi_2016}. Periodic driving can often effectively induce highly non-local hoppings \cite{Tong_2013}.  A periodically driven NLSM  may therefore trigger the formation of multiply linked nodal loops in the resultant Floquet topological phase. We thus consider periodic driving applied to simple, {physically realistic} NLSM systems with only nearest neighbor hoppings.   Remarkably, by a two-step periodic modulation scheme, we obtain Floquet NLSMs with arbitrarily many desired Floquet nodal loops that are linked in controllable, exotic ways.
 With details of realizing simple NLSMs with nearest neighbor hoppings recently outlined in a cold-atom experimental proposal {\cite{DWZhang_2017}}, our scheme
  is already within reach of today's experiments.
 Thus, our work not just reinforces the notion that periodic driving is fruitful in NLSM systems \cite{driven_NL1,driven_NL2,driven_NL3,driven_NL4,Yan_2016,Yan_2017}, but also opens up an avenue towards experimental realization of the most exotic NLSM topological phases to date.   {In addition to offering two theoretical tools to characterize the obtained intricate linkages between the many nodal loops, we also discuss several routes to experimental measurements.}

Nodal lines in 3D systems are protected by certain crystal symmetries such that {band-crossings occur when two constraints are simultaneously satisfied. 
Consider, for example, a two-band dimensionless (with $\hbar=1$) Hamiltonian $H=\bm{h}({\bm k})\cdot\bm{\sigma}=h^x({\bm k})\sigma_x+h^z({\bm k})\sigma_z$, where $\sigma_{x,z}$ are the standard Pauli matrices. The absence of $\sigma_y$ in $H$ reflects a sublattice symmetry, and yields nodal lines as loci of momenta where the coefficients $h^x({\bm k})$ and $h^z({\bm k})$ of $\sigma_{x}$ and $\sigma_{z}$ both vanish.
For a specific and experimentally relevant system, we consider two such NLSM Hamiltonians $H_1$ and $H_2$, with their respective $x$-$z$ components given by
\begin{eqnarray}
H_1:\;\;&&h_1^x=\sin k_z, h_1^z=\cos(k_x+\phi)+\cos k_y-{\mu};\nonumber \label{1_model2}\\
H_2:\;\;&&h_2^x=\sin k_y, h_2^z=\cos(k_x-\phi)+\cos k_z-{\mu},\label{l_model1}
\end{eqnarray}
where $k_x$, $k_y$, and $k_z$ are the three components of the Bloch momentum ${\bm k}$ along different directions.
As proposed in \cite{DWZhang_2017}, $H_1$ or $H_2$ with $\phi=0$
can be realized in a cubic optical lattice with each unit cell having two sublattice sites. The parameter $\phi$ introduced here represents a complex nearest-neighor hopping amplitude along the $x$ direction. This can be effectively generated by a high-frequency rocking linear force along $x$ \cite{Sols2008},  which renormalizes the hopping with a phase.   Fig. \ref{fig1} depicts one nodal loop each for $H_1$ and $H_2$.  In certain parameter regimes [panel (b)], the two loops shown in Fig.~\ref{fig1} appears to be linked.  Nevertheless, this linkage does not represent any new phase, because they arise simply from two independent systems that are simultaneously plotted.
\begin{figure}
\includegraphics[width=0.9\linewidth]{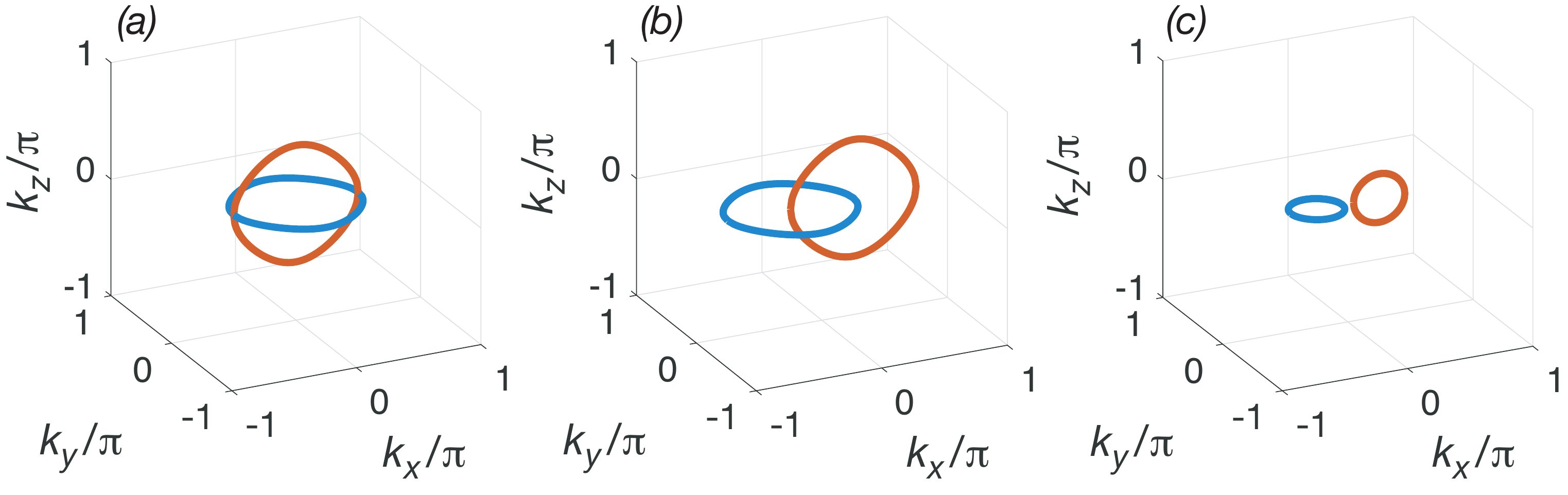}
\caption{Nodal loops of $H_1$ and $H_2$ specified by Eqs. (\ref{l_model1}). (a) $\phi=0$, ${\mu}=1$; (b) $\phi=\pi/4$, ${\mu}=1$ and (c) $\phi=\pi/4$, ${\mu}=1.8$. The blue (red) loop in each panel corresponds to gapless $H_1$ ($H_2$) at $k_z=0$ ($k_y=0$). In (a), the two loops are touching;
in (b), they are linked whereas in (c) they are not linked or touching.}
\label{fig1}
\end{figure}
We now consider a periodically driven {scenario with overall Floquet period $T=T_1+T_2$, such that} $H= H_1$ for duration $T_1$ and $H=H_2$ for duration $T_2$. As $H_1$ and $H_2$ take rather similar forms in Eq.~(\ref{1_model2}), this driving scheme simply requires {easily implementable} periodic switching of the relevant optical lattice potentials along $y$ and $z$, plus a  periodic modulation {of $\phi$, the phase offset }
of the high-frequency field along $x$. The associated {single}-period evolution operator $\hat{U}_T=e^{-i\int_0^TH(t)dt}$ defines an effective Floquet Hamiltonian $H_{\rm eff}$ through $\hat{U}_T=e^{-iH_{\rm eff}T}$.  Floquet eigenstates are obtained from $\hat{U}_T|u\rangle=e^{-i\varepsilon T}|u\rangle$, with $\varepsilon$ the quasienergies, defined up to a multiple of $2\pi/T$ and generally chosen to lie in $(-\pi/T,\pi/T]$.

Through simple algebra~\cite{sup}, $H_{\rm eff}$ is found to also respect sublattice symmetry with no $\sigma_y$ term, and hence also describe a Floquet NLSM.
The closed SU(2) algebra allows us to find explicit conditions for the two Floquet bands of $\hat{U}_T$ to touch.
These conditions are \cite{sup}
\begin{eqnarray}
&&\bm{\hat{h}}_1({\bm k})=-\bm{\hat{h}}_2({\bm k}),~|\bm{h}_1({\bm k})|-|\bm{h}_2({\bm k})|=\Delta_n,~\mathrm{or}\label{con1}\\
&&\bm{\hat{h}}_1({\bm k})=\bm{\hat{h}}_2({\bm k}),~|\bm{h}_1({\bm k})|+|\bm{h}_2({\bm k})|=\Delta_n,\label{con2}
\end{eqnarray}
with $\Delta_n= 2n\pi/T$, $n$ an arbitrary integer, and $\bm{\hat{h}}_i({\bm k})=\bm{h}_i({\bm k})/|\bm{h}_i({\bm k})|$. Here we set $T_1=T_2=T/2$ without loss of generality.  The two band-touching conditions in either Eq.~(\ref{con1}) or Eq.~(\ref{con2}) define 1D band-touching lines as nodal lines.  The freedom in choosing different $n$ stems from the fact that $\varepsilon= 2m\pi/T$ [$\varepsilon=(2m+1)\pi/T$] with different integers $m$  are all physically equivalent to $\varepsilon= 0$ [$\varepsilon=\pi/T$].
It is this qualitative insight that suggests the possibility of multiple linked nodal loops at $\varepsilon=0$ and $\pi/T$.
Other than the solutions from Eqs.~(\ref{con1}) and (\ref{con2}), nodal lines also appear at $T_1|\bm{h}_1({\bm k})|=n_1 \pi$
and $T_2|\bm{h}_2({\bm k})|=n_2 \pi$, with $n_1$ and $n_2$ also integers.  These extra solutions are of no interest because no extra linked nodal loops are observed, and so they are not discussed in our following systematic analysis.
Thus {Eqs.~(\ref{con1}) and (\ref{con2}) are the main defining equations for our {Floquet system} with exotic nodal lines.

%

It can now be hoped that given a sufficiently large $T$, the above nodal line solutions yield many nodal loops, each one determined by a particular choice of $n$ .
As an encouraging computational example, Fig.~\ref{fig2} shows that 3, 4 and 5 linked nodal loops in the Floquet band structure can indeed be generated, along with other nodal lines.
\begin{figure}
\includegraphics[width=1\linewidth]{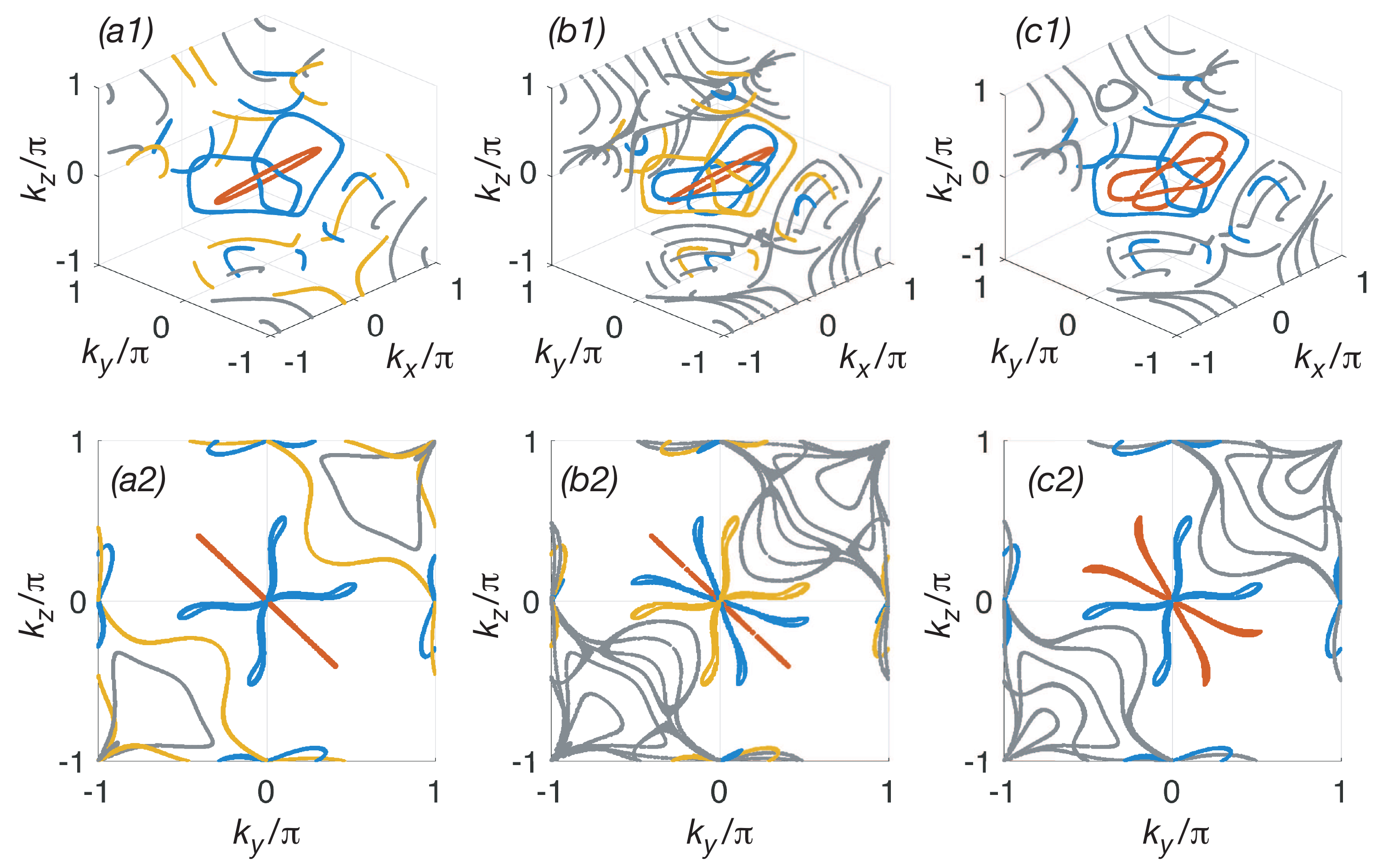}
\caption{Multiply linked nodal loops from Eqs.~(\ref{con1}) and (\ref{con2}) for the Hamiltonian of Eq.~(\ref{l_model1}) with ${\mu}=1$ and $\phi=\pi/4$. (a) three linked nodal loops at $\varepsilon=0$ with $T=3\pi$, (b) five linked nodal loops at $\varepsilon=0$ with $T=5\pi$. Red, blue and yellow lines are for $n=0$, $n=\pm 2$ and $n=\pm4$ respectively, and gray lines for larger even $n$. (c) four linked nodal loops at $\varepsilon=\pi/T$ with $T=4.5\pi$. Red and blue lines are for $n=\pm 1$ and $n=\pm 3$ respectively. Gray lines represent nodal lines obtained from larger odd $n$. In either case, gray lines do not produce linked nodal loops.  (a2) to (c2) are the side views of (a1) to (c1).}\label{fig2}
\end{figure}

To understand the appearance and shapes of the possible nodal lines/loops arising from Eqs.~(\ref{con1}) and (\ref{con2}), we treat $\Delta_n$ as a single adjustable parameter, tunable via experimentally controlling $T$. We first consider the simplifying regime of small $\Delta_n$ (i.e. small $2\pi n/T$), which ensures that the nodal solutions emerge only from Eq.~(\ref{con1}) but not from Eq. (\ref{con2}). We next define
\begin{equation}
\alpha_j=\frac{h_1^j({\bm k})}{\sqrt{{|\bm{h}_1({\bm k})|}}},\  \beta_j=\frac{h_2^j({\bm k})}{\sqrt{{|\bm{h}_2({\bm k})|}}},
\end{equation}
with $j=x,z$;
and construct the complex variables
\begin{equation}
z(\bm{k})=\alpha_x+i\beta_z,~~w(\bm{k})=\beta_x+i \alpha_z.
\end{equation}
Eq.~(\ref{con1}) is then found to be equivalent to the compact condition \cite{sup}.
\begin{equation}
g_n(\bm{k})\equiv z(\bm{k})+ \sqrt{w^2(\bm{k})+\Delta_n}=0. \label{gdelta}
\end{equation}
That is, nodal solutions are found when the real and imaginary parts of $g_n(\bm{k})$ separately vanish.
Equation~(\ref{gdelta}) also leads to a mathematically transparent way of computing} the linkage of nodal loops given by $g_n(\bm{k})$ with different $n$. Indeed, as shown rigorously in \cite{sup}, we find that any two such nodal loops are necessarily linked with unit linking number, provided that the original nodal loops of $H_1$ and $H_2$ are linked.

{In the limit of diverging period $T$ or vanishing small $|\Delta_n|$, the Floquet quasienergies coalesce and the defining equations Eqs.~(\ref{con1}) and~(\ref{con2}) can} accommodate a large number of nodal solutions corresponding to different $n$, hence hosting as many linked nodal loops as we wish. For a finite $T$, however, one would wonder how many pairwise linked loops can be generated, each corresponding to different $n$.
With details elaborated in Supplementary Material~\cite{sup}, we summarize our main {observations/results} in the following.

For a given finite $T$, as the chosen value of $|n|$ continues to increase beyond a critical value, nodal loops given by $g_{n}(\bm{k})=0 $ and $g_{-n}(\bm{k})=0 $ start to split into different segments.  Interestingly, the end points of these segmented solutions are exactly the solutions
  of Eq.~(\ref{con2}). This is also when Eq.~(\ref{con2}) starts to yield nodal line solutions. Fig.~\ref{fig3}(a)-(c) illustrates an example of how a pair of linked nodal loops, initially simultaneous solutions to  Eq.~(\ref{con1}) at a chosen $|n|$, start to split as $T$ decreases, and are connected by the solutions to Eq.~(\ref{con2}). Eventually, at sufficiently small $T$, the nodal loci are all taken over by the solutions to Eq.~(\ref{con2}) with the same $n=|n|$.
	
The nodal solutions to Eq.~(\ref{con2}) associated with a single chosen $n=|n|$ can merge or split as we tune system parameters such as $T$. This differs from that solutions to $g_{n}(\bm{k})=0 $ and $g_{-n}(\bm{k})=0$, which cannot intersect. In our model system [Eq.~(\ref{l_model1})], these transitions are found to occur when different segments of the nodal line solutions touch at
${\bm k_0}=(0,0,0)$
 in the Brillouin zone \cite{sup}, when $T$ crosses the critical value {$T=\pi |n|/(\cos{\phi}+1-{\mu})$} [see Fig.~\ref{fig3}(d)-(f)].
In other words,  for a given $n=|n|$, as $T$ is increased above the critical value, two nodal lines become linked loops; and for a fixed $T$,  every even (odd) $n$ satisfying $(\cos\phi+1-\mu)T/\pi>n>0$ yields two linked nodal loops at $\varepsilon=0$ ($\varepsilon=\pi/T$).
The solution with $n=0$ represents a single nodal loop and further increases the number of pairwise linked loops at $\varepsilon=0$  by one.
As a final side observation, Eqs.~(\ref{con1}) and (\ref{con2}) also give some unlinked nodal lines which may wrap around the whole BZ \cite{footnote}.


\begin{figure}
\includegraphics[width=0.99\linewidth]{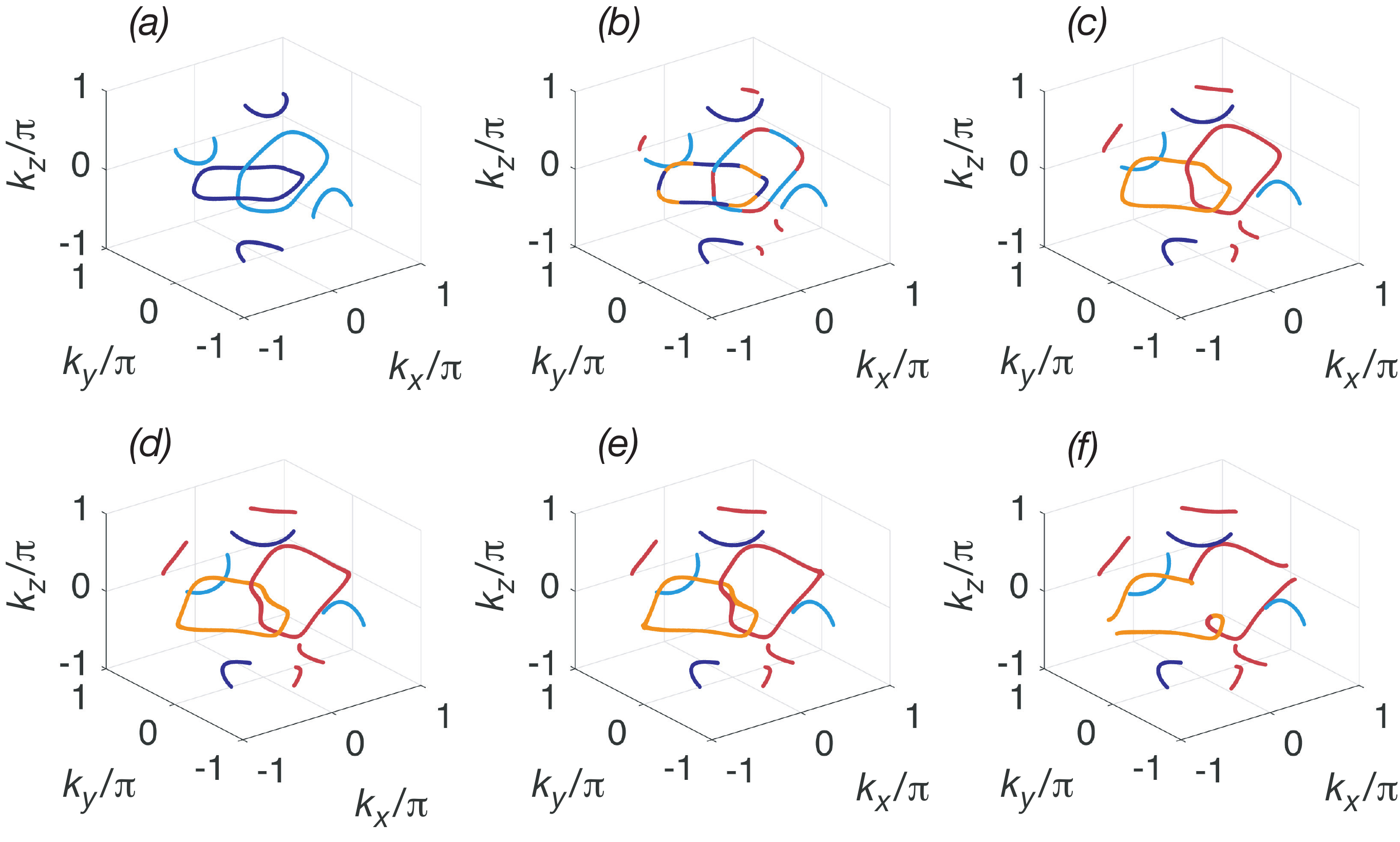}
\caption{{Transition of nodal loci of} Eqs. (\ref{con1}) and (\ref{con2}) at a single fixed value of $|n|$ and increasing $\Delta_n$ (decreasing $T$). The parameters are ${\mu}=1$, $\phi=\pi/4$ and (a) $\Delta_{n}=0.9$, (b) $\Delta_{n}=1.1$, (c) $\Delta_{n}=1.3$, (d) $\Delta_{n}=1.4$ (e) $\Delta_{n}=\sqrt{2}$, and (f) $\Delta_n=1.5$. Red {and orange} curves are the solutions of Eqs. (\ref{con2}) (different colors are to guide eyes only).   Light and dark blue curves are respectively obtained from $g_{n}({\bm k})=0$ and $g_{-n}({\bm k})=0$.
Panels (a) to (c) illustrate how the pairwise link initially given by Eq. (\ref{con1}) are taken over by solutions to Eq. (\ref{con2}). Panels (d) to (f) illustrate the eventual breaking of the pairwise link given by Eq. (\ref{con2}) as $\Delta_n$ continues to increase.
} \label{fig3}
\end{figure}

{Next we discuss the topological invariants characterizing different linkages. We first} introduce the winding number which reveals {the transition at $T=\pi |n|/(\cos{\phi}+1-{\mu})$}. 
{The winding number for $H_{\rm eff}=h_x\sigma_x+h_z\sigma_z$ along a closed trajectory $c$ parametrized by $\theta$ is}
\begin{eqnarray}
\nu=\frac{1}{2\pi}\oint_{c}d\theta \frac{h_x\partial_\theta h_z-h_z\partial_\theta h_x}{h_x^2+h_z^2}\label{winding}
\end{eqnarray}
\text{One obtains} $\nu=\pm1$, which indicates a $\pi$ Berry phase, if the closed trajectory is linked with a nodal loop. 
To detect the pairwise linkage between two nodal loops, one can choose two closed trajectories associated with the eventual touching point $\bm k_0=0$, as { detailed in  [\onlinecite{sup}]. Specifically, as the Floquet period $T$ changes, the merging or splitting of two nodal loop solutions of Eq.~(\ref{con2}) (for the same $n$) are well-captured by jumps in the windings of two chosen differently-oriented small circle trajectories $c_1:~k_x=r\sin{\theta},~k_y=k_z=r\cos{\theta}$, and
$c_2:~k_x=r\sin{\theta},~k_y=-k_z=r\cos{\theta}$. In the $\nu_1,\nu_2$ plots in Fig.~\ref{fig4}}, we can unambiguously identify the critical period $T$ by examining when the $\pm 2$ winding collapses as the radius $r\rightarrow 0$ .}

\begin{figure}
\includegraphics[width=1\linewidth]{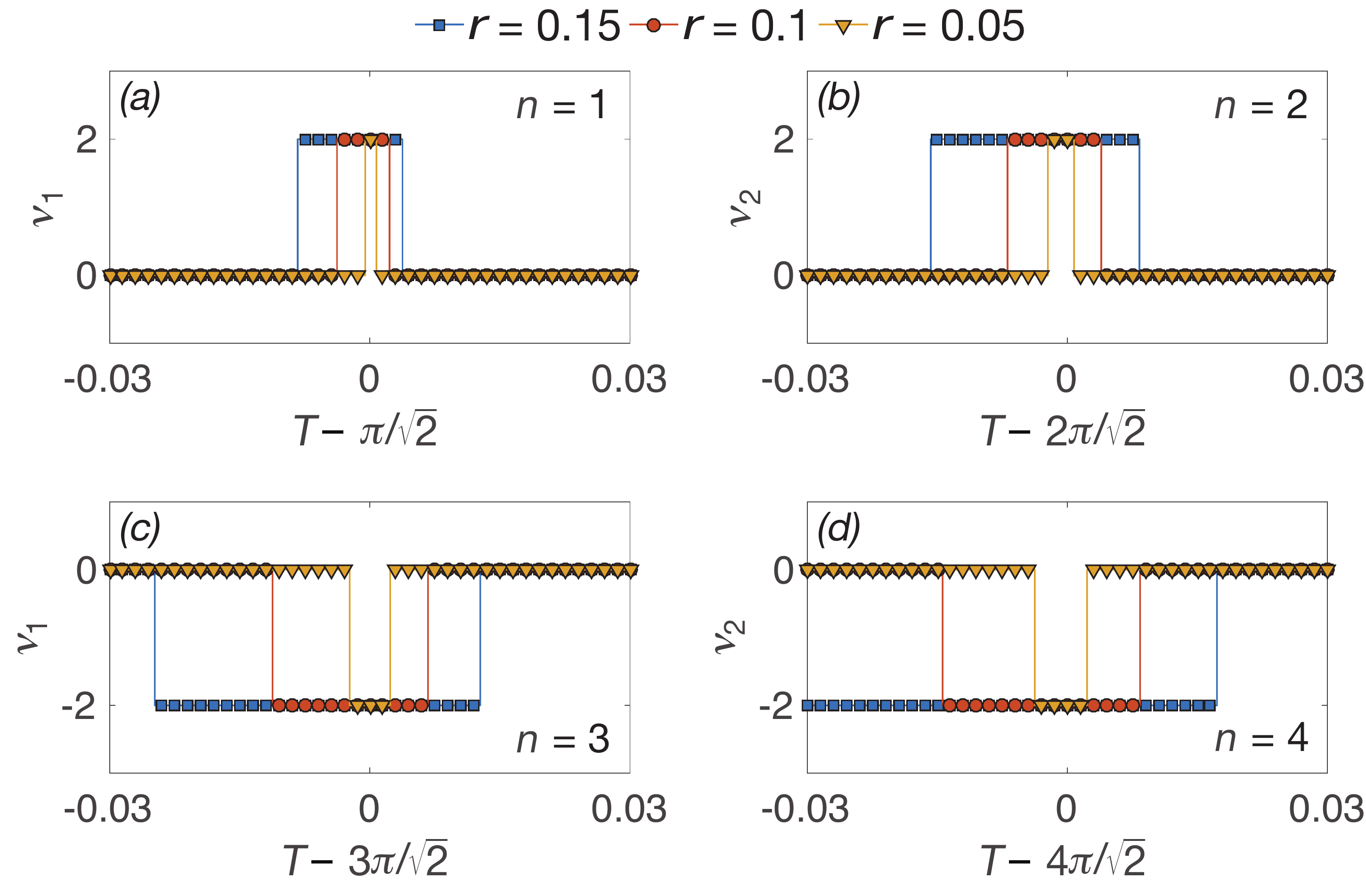}
\caption{The winding numbers $\nu_{1,2}$ {against} $T$, with $\mu=1$, $\phi=\pi/4$, {demonstrating a critical $T$ of} $T=\pi|n|/\sqrt{2}$. (a) to (d) show different regimes of $T$, corresponding to the transition of the $n$-th pair of Hopf-link respectively {(as labeled in each panel)}. Different colors and marks correspond to different value of $r$, the ``radius" of the two trajectories mentioned in the text.
 (a) and (c) only show the value of $\nu_1$, while $\nu_2$ is zero in these regimes; For (b) and (d), $\nu_1$ is zero and we only show $\nu_2$.
}\label{fig4}
\end{figure}

{To further characterize the complicated topology of the multiple nodal linkages, we need simultaneous winding information from multiple closed trajectories. Collectively, they yield the \emph{Alexander polynomial} $A(t)$, an established topological invariant for discerning superficially similar knots or multiply-linked loops in 3D space (Fig.~\ref{fig5}).
For instance, $A(t)$ of the 3 linked loops in Fig.~2a is $2-(t+t^{-1})=-(1-t)^2/t$, i.e. equivalent to the standard 3-link $L6n1$, but not the two Brunnian links $L6a4$ with $A(t)=(t-1)^4/t^2$ or $L10a140$ with $A(t)=(t-1)^4(t^2+1)^2/t^4$, both with trivial pairwise links.}  
{Computing $A(t)$ of a multiple link requires the linking numbers of the various homology loops of its Seifert surface. This is achieved either via its Seifert matrix, or by systematic redrawing of the multiple link into the closure of its representative braid (Fig.~\ref{fig5})~\cite{sup}.}

\begin{table}[H]
\begin{minipage}{\linewidth}
\centering
\renewcommand{\arraystretch}{2}
\begin{tabular}{|l|l|l|}\hline
 \bf{Link} & \bf{Corresponding Braid} & \bf{$A(t)$} \\    \hline
\bf{Hopf} &\ $\tau_1\tau_1$ or $\bar\tau_1\bar\tau_1$ &\ $\frac{\pm(t-1)}{\sqrt{t}}$ \\ \hline
\bf{3-link} &\ $\bar\tau_2\tau_1\tau_2\tau_2\tau_1\bar\tau_2$ &\ $\frac{-(1-t)^2}{t}$ \\ \hline
\bf{4-link} &\ $\tau_2\tau_1\bar\tau_2\bar\tau_1\tau_3\bar\tau_2\tau_1\tau_2\bar\tau_1\tau_2\tau_3\tau_1\tau_2\bar\tau_1$ &\ $0$ \\ \hline
\bf{5-link} &\ $\bar\tau_1\bar\tau_2\tau_3\tau_2\tau_4\tau_1\bar\tau_2\bar\tau_1\tau_3\bar\tau_2\bar\tau_4\tau_1\tau_2\bar\tau_3\tau_2\tau_1\bar\tau_2\tau_3\tau_2\tau_4\tau_3\bar\tau_2$ &\ $\frac{(1-t)^4}{t^2}$ \\ \hline
\end{tabular}
\end{minipage}
\end{table}
\begin{figure}[H]
\includegraphics[width=1.04\linewidth]{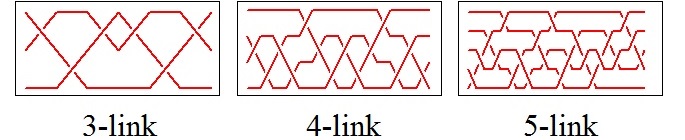}
\caption{{Topological characterization of the multiply-linked loops of Fig. 2, which are also closures of (non-unique) representative braids. $\tau_i$ ($\bar\tau_i$) denotes an overcrossing(undercrossing) of strand $i$ with $i+1$.} 
Interestingly, the 4-link in Fig.~2(c) has Alexander polynomial $A(t)=0$, {unlike many other 4-links with 12 crossings.}}
\label{fig5}
\end{figure}

{Experimentally, signatures of multiply linked nodal loops may be observed by the following complementary approaches.
 {First, Berry phase winding numbers $\nu_1$ and $\nu_2$ and even detailed pseudospin structure~\cite{kohl2005fermionic,alba2011seeing,hauke2014tomography,LMDuan} can be directly measured via Bloch state tomography, as demonstrated }in both static and driven cold atom systems~\cite{flaschner2016experimental,li2016bloch} along arbitrary momentum space trajectories. {Second, peculiarly linked nodal loops also yields} measurable non-linear atomic cloud center-of-mass responses~\cite{price2016measurement,lohse2018exploring} with characteristic frequency multiplication properties in specific directions~\cite{lee2015negative}. Third, their signature nodal Seifert surfaces are physically manifested 
{as surface drumhead states with very large density of states. These can be momentum-resolved to reconstruct the nodal topological structure, or }
detected via a cold-atom variant of a quasiparticle interference experiment~\cite{biderang2018drumhead} or two-terminal conductance measurements \cite{transport_coldatoms1,transport_coldatoms2}, {as further elaborated in the Supplementary Material~\cite{sup}. Beyond cold-atoms, our results can be physically simulated with a controllable parameter space in lieu of 3D momentum space.} For instance, Hopf-link semimetals (simpler than what we have here) can be realized with a superconducting transmon qubit embedded in a 3D aluminium cavity \cite{SC_circuits}.  An analogous experiment setup with some system parameters periodically driven should be useful to simulate the formation of multiple linked nodal loops. For example, observation of Berry phase jumps on such a platform can then confirm the jumps of the winding numbers illustrated in Fig. \ref{fig4}.
 }



To conclude, we have introduced an elegant and experimentally feasible approach to having arbitrarily many nodal loops linked in exotic, desired ways from only nearest neighbor hoppings, thereby providing another promising example of Floquet engineering.
We also hope to have provided an elegant framework for illuminating the intricate connections between topology and complex analytic properties~\cite{Ezawa_Hopf,lee2015free,lee2016band,bode2017knotted}.
{Although we have specifically focused on one physical Hamiltonian recently shown to be experimentally accessible, variations will most likely lead to qualitatively different topological linkages.}  Extending our two-step quenching to multi-step driving protocols may also result in even more complicated nodal line topologies, e.g. the Borromean rings and the Trefoil knot in Floquet bands.

\vspace{1cm}
\begin{acknowledgements}
\emph{Acknowledgements.}-- J. Gong is supported by Singapore Ministry of Education Academic Research Fund Tier I (WBS No.~R-144-000-353-112) and by the Singapore NRF grant No.~NRF-NRFI2017-04 (WBS No.~R-144-000-378-281). L. L. would like to thank Tian-Shi Xiong for helpful discussions.
\end{acknowledgements}

\clearpage
\onecolumngrid
\setcounter{equation}{0}
\setcounter{figure}{0}
\setcounter{table}{0}
\makeatletter
\renewcommand{\theequation}{S\arabic{equation}}
\renewcommand{\thefigure}{S\arabic{figure}}
\begin{center}
\textbf{Supplementary Material: Realistic Floquet semimetal with exotic topological linkages between arbitrarily many nodal loops }\\[5pt]
\vspace{0.1cm}
\begin{quote}
\end{quote}
\end{center}

\setcounter{equation}{0}
\setcounter{figure}{0}
\setcounter{table}{0}
\setcounter{page}{1}
\setcounter{section}{0}
\makeatletter

\maketitle
This supplementary material contains three main sections. In the first section we introduce the Alexander polynomial as a topological invariant to characterize topologically different linkage among the nodal loops. In the second section, we give a detailed analysis of the nodal solutions in our Floquet system.  There we first derive the general solutions for two-step quenching systems, and then simplify them to the zeros of a series of complex functions $g_n({\bm k})$.  Next we investigate the linkages of the nodal solutions and their transitions. At the end of this section we introduce a winding number to characterize the possible touching points of the nodal lines.   {We devote Section III to experimental characterization of the linked nodal loops discovered in our work}.

\section{Section I: Calculation of the topological invariants of the nodal links}
\label{sec:alexander}
In this work, we have considered 1D nodal sets (which we here denote as $l$) embedded in a 3D BZ. The topological classification of such nodal sets is extremely rich, encompassing an infinite set of topologically inequivalent nodal knots and links. Unlike $\mathbb{Z}_2$ or Chern bundles in 2D which are completely characterized by a sign or an integer, there is no single topological quantity that provides an unambiguous 1-to-1 classification of a system of linked nodal rings. While there exist several different types of polynomial topological invariants for knot and links, none of them can individually distinguish all possible inequivalent knots or links - indeed for each type of polynomial invariant, two links corresponding to different polynomials are definitely not equivalent, but two links with the same polynomial may or may not be equivalent. As such, we shall compute two of the most convenient and discerning topological invariants for our linked nodal rings, the Alexander polynomial and the link signature. These are two of the oldest established knot invariants, with the Alexander polynomial the only known knot polynomial until 1984, when the Jones polynomial was discovered via a very different perspective\footnote{The Alexander-Conway polynomial $A'(t)$, which was developed in 1969, is equivalent to the Alexander polynomial $A(t)$ via $A'(t^{1/2}-t^{-1/2})=A(t)$.}.

Both of these invariants can be most conveniently computed via the Seifert matrix $S$. To explain the computation procedure, we first define the Seifert surface $\Sigma$ of a given knot or link $l$ as an orientable surface whose boundary is $l$ (Fig~\ref{fig:Seifertview}). Such a surface can always be constructed by "interpolating" across the space between different parts $l$, and must be orientable because its boundary curves (the knots or links) are closed, orientable loops.

Note that a Seifert surface $\Sigma$ of a knot or link is highly non-unique, because there are many ways to embed the knot/link in 3D space. To construct topological invariants out of it, we need a systematic way to extract its intrinsic winding properties. This can be done by studying the linkings of the homology generators (loops) of $\Sigma$. However, since closed loops on a surface generically intersect, we need to define an infinitesimally "lifted" Seifert surface~\cite{murasugi}
$\Sigma^\#$ that is slightly "above" the original surface $\Sigma$ at every point, which is always consistently defined since $\Sigma$ is orientable. With that, we can arbitrarily choose a basis set of 1st homology generators (loops) $\alpha_i$, $i=1,...,r$ that generates the 1st homology group of $\Sigma$, with corresponding lifted loops $\alpha^\#_i$ in $\Sigma^\#$. For instance, if $\Sigma$ (or $\Sigma^\#$) is homeomorphic to a disk with $r$ holes, the $r$ homology loops can be chosen to encircle each of the $r$ holes.

With a basis $\{\alpha_i\},\{\alpha^\#_i\}$ defined, we can compute the central quantity, the $r\times r$ Seifert matrix $S$ of linking numbers:
\begin{equation}
S_{ij}=\text{Lk}(\alpha_i,\alpha_j^\#)=\frac1{4\pi}\int_{\alpha_i}d\vec x \int_{\alpha^\#_j}d\vec y \frac{(\vec x-\vec y)\cdot (d\vec x\times d\vec y)}{|\vec x-\vec y|^3}.
\end{equation}
Geometrically, the linking number of two curves is the total winding of one curve about the other, and is only well-defined if the two curves never touch (Hence the necessity of the lifting). While $S$ depends on the particular presentation of the knot/link, the combination of $S$ and its transpose yields two topological invariants:
\begin{enumerate}
\item The Alexander polynomial invariant $A(t)$ given by
\begin{equation}
A(t)=t^{-r/2}\text{Det}(S-tS^T),
\end{equation}
\item The signature $\sigma$ of the knot/link, which is the difference between the number of positive and negative eigenvalues of $S+S^T$.
\end{enumerate}
In the above, we see that the topological invariants $A(t)$ and $\sigma$ rely crucially on the linking number $\text{Lk}(\alpha_i,\alpha_j^\#)$ of the homology loops. Although also a topological invariant, the linking number is admittedly a very simple one that distinguishes different knots/links with limited success. But when applied to every pair of homology loops of a Seifert surface of a given knot/link, it yields much more discerning topological invariants $A(t)$ and $\sigma$.

\subsection{Seifert matrix, Alexander polynomial and signature of quenched nodal links}
For illustration purposes, we detail the computation of the linked nodal loops in Fig. 2a, where a red nodal ring from the $n=0$ sector ($ \bold {\hat h_1}=-\bold {\hat h_2}$ and $|\bold {\hat h_1}|=|\bold {\hat h_2}|$) links nontrivially with a blue Hopf link from the $n=2$ sector ($ \bold {\hat h_1}=\bold {\hat h_2}$ and $|\bold {\hat h_1}|+|\bold {\hat h_2}|=\Delta=2\pi n/T$ with $T=3\pi$). The resultant nodal structure $l$ contains 3 linked nodal loops, and we shall compute its topological invariants to identify it among a few other topologically inequivalent 3-component links.

\begin{figure}[H]
\begin{minipage}{\linewidth}
\subfloat[]{\includegraphics[width=0.24\linewidth]{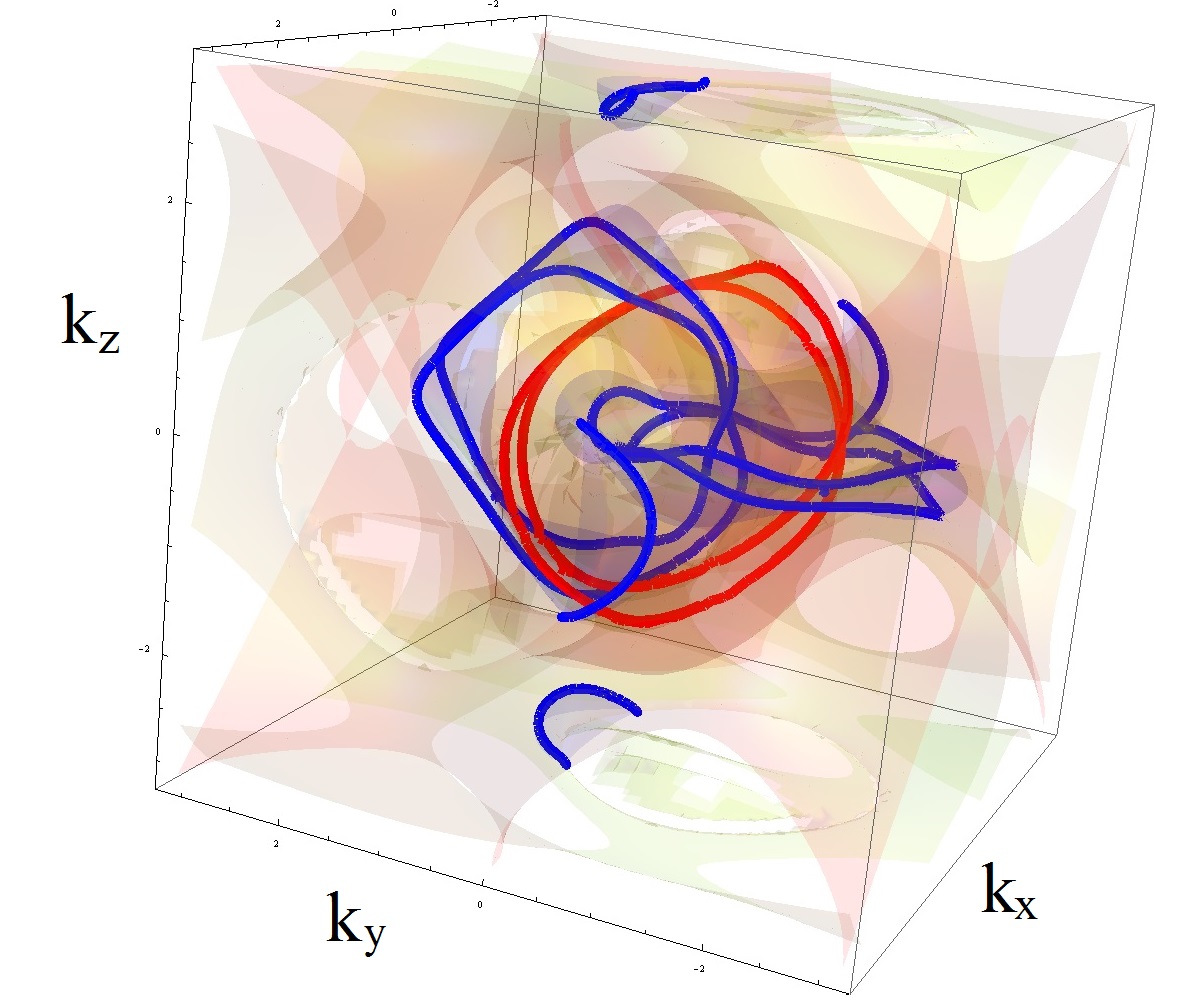}\label{fig:Seifert_lifted}}
\subfloat[]{\includegraphics[width=0.16\linewidth]{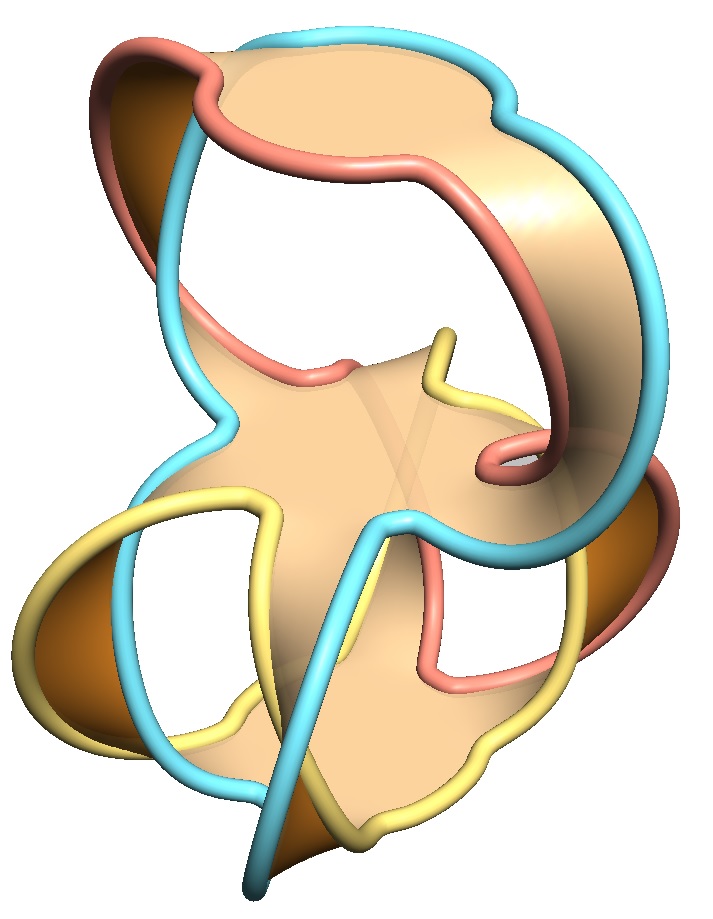}\label{fig:Seifertview}}
\subfloat[]{\includegraphics[width=0.58\linewidth]{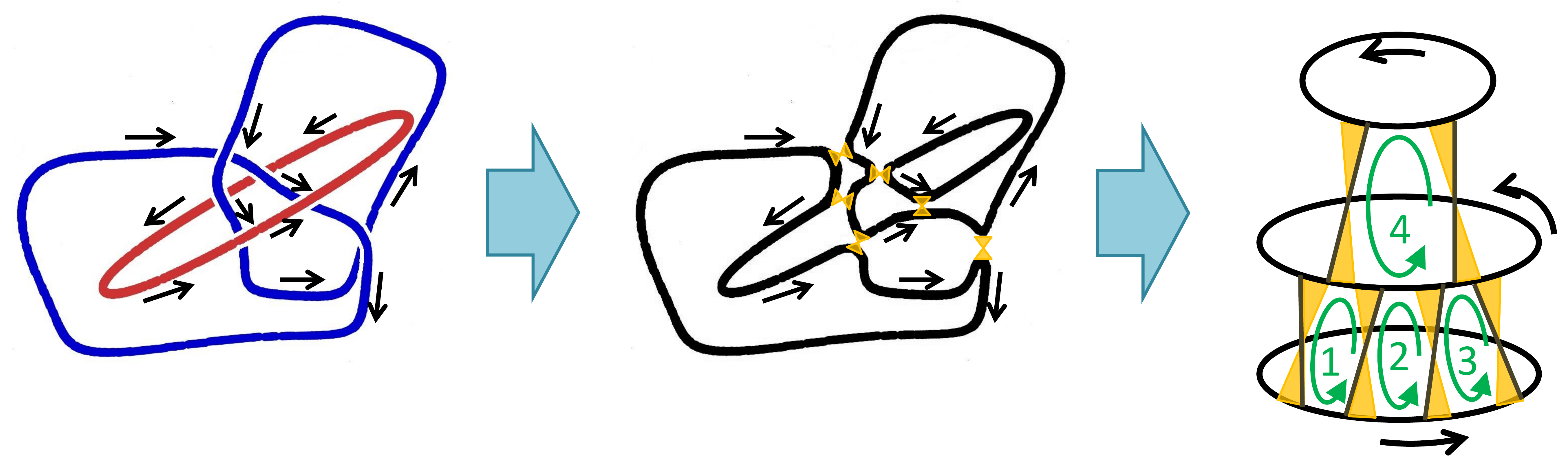}\label{fig:Seifert_draw}}
\end{minipage}
\caption{a) the link $l$ defined by the $n=0$ and $n=2$ sectors of Fig. 2a, together with its lifted $l^\#$. There is no need to distinguish between $l$ and $l^\#$, which are infinitesimally shifted (here the shift was computed with $\Delta$ differing by a large $\approx 5\%$ for clarity's sake.) b) An aesthetic rendering of $l$ and its Seifert surface $\Sigma$ with Seifertview. c) The procedure for mathematically constructing the Seifert surface and the homology generators (green, labeled $i=1,2,3,4$). With different choices of orientations, we may obtain different Seifert surfaces and hence its homology generators and Seifert matrix. However, the topological invariants hence obtained remain unaffected. }
\label{fig:Seifert2}
\end{figure}

The first step is to generate the "lifted" nodal link $l^\#$ from our original link $l$. One way, which is by no means unique, is to simply take $l^\#$ as the link resulting from a slightly shifted $\Delta$ (Fig. \ref{fig:Seifert_lifted}). This construction of $l^\#$ is valid as long as $l$ and $l^\#$ never intersect. In general, this is guaranteed if $l^\#$ is always shifted relative to $l$ along one of the two level surfaces whose intersection define $l$, as long as the Jacobian of the mapping never vanishes.

Next, we construct the homology generators of the Seifert surface $\Sigma$ from $l$,and its lifted counterpart $\Sigma^\#$ from $l^\#$. A heuristically convenient way proceeds as illustrated in Fig. \ref{fig:Seifert_draw}. Considering $l$ and $l^\#$ separately, we first choose an orientation of the link diagram as in the leftmost panel. Next, we represent each crossing with a ``twist" structure, as indicated in yellow in the middle panel. The twists are orientated such that the oriented links become closed loops consistent with the chosen orientation. Finally, we redraw the resultant structure in a standard "wedding cake" style as in the rightmost panel, where the closed loops now bound each flat circle of the ``cake", with the twist operators connecting the various circles. We have constructed a bona-fide Seifert surface for $l$, with the boundaries of the ``wedding cake" clearly corresponding to the original $l$, and the twist operators taking into account of the structure of its crossings. In this standard form, it is easy to define a homology basis (green) consisting of the independent closed loops traversing the twist operators. From this wedding cake, it is also not difficult to deduce the a representative braid for the knot/link by identifying the twist operators as braid operations, and the layers of the cake as the strands~\cite{collins2016algorithm}.

To obtain the Seifert matrix, we compute the linking numbers of each of the homology generators $\alpha_i$ (green) of $\Sigma$ with the each of the lifted $\alpha_j^\#$ of $\Sigma^\#$ (not shown). This requires some imagination, but heuristically, the linking number of a loop with its lifted version $\text{Lk}(\alpha_i,\alpha_i^\#)$ is equal to half the total winding number accrued in the twist operators it traversed. Also, the $\text{Lk}(\alpha_i,\alpha_j^\#)$ for $i\neq j$ depends only on the sign of the twist operator "between" them. With some care, we obtain the Seifert matrix of our 3-component link $l$ in Fig. \ref{fig:Seifert2} as
\begin{equation}
S_{3-link}=\left(\begin{matrix}
0 & -1 & 0 & 1 \\
0 & 1 & -1 & 0 \\
0 & 0 & 0 & -1 \\
0 & 0 & 0 & 1 \end{matrix}\right)
\end{equation}
with the basis defined as in the rightmost panel of Fig. \ref{fig:Seifert_draw}. We hence obtain its Alexander polynomial invariant
\begin{equation}
A_{3-link}(t)=t^{-4/2}\text{Det}(S-tS^T)=t^{-2}(-t+2t^2-t^3)=2-\left(t+\frac1{t}\right).
\end{equation}
Since $S+S^T$ has doubly degenerate eigenvalues $1\pm \sqrt{3}$, the signature of $l$ is zero. Note that we will have obtained a zero $A(t)$ had we included the other unlinked loops in the full nodal set in Fig 2a, because the Alexander polynomial vanishes trivially when two or more unlinked components are present. $A_l(t)$ identifies our nodal link with the standard nomenclature $L6n1$, and distinguishes it with the Borromean ring with $A(t)=\left(t^2+\frac1{t^2}\right)-4\left(t+\frac1{t}\right)+6$, or other 3-component links with $A(t)=\left(t^2+\frac1{t^2}\right)-\left(t+\frac1{t}\right)$ or $A(t)=-\left(t^2+\frac1{t^2}\right)+3\left(t+\frac1{t}\right)-4$.

A useful property of the Alexander polynomial allows us to write it down for an arbitrarily linked nodal chain (without 4-fold degenerate intersections). Namely, $A(t)=A_1(t)A_2(t)$ if $A(t)$ corresponds to the connected sum of two knots/links corresponding to $A_1(t)$ and $A_2(t)$. Since a nodal chain with $N$ loops is just the connected sum of $N-1$ Hopf links, each with a $1\times 1$ Seifert matrix $S=\pm 1$ corresponding to its single homology loop, the Alexander polynomial of a nodal chain must be given by $A_{\text{chain}}(t)=\pm \left(\frac1{\sqrt{t}}-\sqrt{t}\right)^{N-1}$, with the $\pm$ sign chosen according to the product of the chiralities of each of its constituent links.

\subsubsection{Seifert matrix for the 4-link}

For the 4-link in Fig. 2, a similar (though more tedious) analysis as above gives the Seifert matrix
\begin{equation}
S_{4-link}=\left(\begin{array}{*{11}c}
0& 0&  -1  &   0& 0& 0& 0& 0& 0& 0& 0 \\
  -1& 0& 1  &  -1& 0& 0& 0& 0& 0& 0& 0\\
 0& 0& 1& 0& 1  &  -1& 0& 0& 0& 0& 0\\
 0& 0  &  -1& 0& 0& 1& 0& 0& 0& 0& 0\\
 0& 0& 0& 0  &  -1& 0& 0& 0& 0& 0& 0\\
 0& 0& 0& 0& 0& 0& 0& 0& 0& 0& 0\\
 0& 0& 0& 1& 0  &  -1& 0& 1  &  -1& 0& 0\\
 0& 0& 0& 0& 0& 1& 0  &  -1& 0& 0& 0\\
 0& 0& 0& 0& 0& 0& 0  &  -1& 0& 1& 0\\
 0& 0& 0& 0  &  -1& 0& 0& 1& 0   & -1& 0\\
 0& 0& 0& 0& 0& 0& 0& 0& 1  &  -1& 0
\end{array}\right)
\end{equation}
It is an $11$-by-$11$ matrix since there are $11$ homology generators, as evident in its braid diagram in Fig. 5. Interestingly, its Alexander polynomial $A_{4-link}(t)=t^{-11/2}\text{Det}(S-tS^{T})=0$ vanishes despite not being a trivial link, unlike other inequivalent 4-links.

\subsubsection{Seifert matrix for the 5-link}

For the 5-link in Fig. 2, further computation gives the Seifert matrix
\begin{equation}
 {S_{5-link}}=\left(\begin{array}{*{18}c}
0 & 0 & 0 & 1 & 0 & 0 & 0 & 0 & 0 & 0 & 0 & 0 & 0 & 0 & 0 & 0 & 0 & 0 \\
  0 & 0 & 1 & 0 & 0 & 0 & 0 & 0 & 0 & 0 & 0 & 0 & 0 & 0 & 0 & 0 & 0 & 0\\
  0 & 0    & -1 & 0 & 1 & 0 & 0 & 0 & 0 & 0 & 0 & 0 & 0 & 0 & 0 & 0 & 0 & 0\\
  0 & 1 & 0 & 0 & 0 & 0   & -1 & 0 & 0 & 0 & 0 & 0 & 0 & 0 & 0 & 0 & 0 & 0\\
  0 & 0 & 0 & 0 & 0 & 0 & 0 & 0 & 0 & 0    & -1 & 0 & 0 & 0 & 0 & 0 & 0 & 0\\
  1 & 0 & 0    & -1 & 0 & 0 & 1    & -1 & 0 & 0 & 0 & 0 & 0 & 0 & 0 & 0 & 0 & 0\\
  0 & 0    & -1 & 0 & 0 & 0 & 1 & 0 & 1    & -1 & 0 & 0 & 0 & 0 & 0 & 0 & 0 & 0\\
  0 & 0 & 0 & 0 & 0 & 0    & -1 & 0 & 0 & 1 & 0 & 0 & 0 & 0 & 0 & 0 & 0 & 0\\
  0 & 0 & 1 & 0    & -1 & 0 & 0 & 0 & 0 & 0 & 1 & 0 & 0    & -1 & 0 & 0 & 0 & 0\\
  0 & 0 & 0 & 0 & 0 & 0 & 0 & 0 & 0 & 0 & 0 & 0 & 0 & 0 & 0 & 0 & 0 & 0\\
  0 & 0 & 0 & 0 & 0 & 0 & 0 & 0 & 0 & 0 & 0 & 0 & 0 & 0 & 0 & 0 & 0 & 0\\
  0 & 0 & 0 & 0 & 0 & 0 & 0 & 1 & 0    & -1 & 0    & -1 & 0 & 0 & 1 & 0 & 0 & 0\\
  0 & 0 & 0 & 0 & 0 & 0 & 0 & 0    & -1 & 1 & 0 & 0    & -1 & 1 & 0 & 0 & 0 & 0\\
  0 & 0 & 0 & 0 & 0 & 0 & 0 & 0 & 0 & 0 & 0 & 0 & 0 & 0 & 0 & 0 & 0 & 0\\
  0 & 0 & 0 & 0 & 0 & 0 & 0 & 0 & 0 & 0 & 0 & 0 & 1 & 0 & 0    & -1 & 0 & 0\\
  0 & 0 & 0 & 0 & 0 & 0 & 0 & 0 & 0 & 0 & 0 & 0 & 0    & -1 & 0 & 0 & 1 & 0\\
  0 & 0 & 0 & 0 & 0 & 0 & 0 & 0 & 0 & 0    & -1 & 0 & 0 & 1 & 0 & 0    & -1 & 0\\
  0 & 0 & 0 & 0 & 0 & 0 & 0 & 0 & 0 & 0 & 0 & 0 & 0 & 0 & 0 & 1    & -1 & 0\\
\end{array}\right)
\end{equation}
It is an $18$-by-$18$ matrix now, with $18$ homology generators for the Seifert surface traced out from Fig. 2. While the Seifert surface and its (number of) homology generators are not uniquely specified,  the Alexander polynomial is unambiguously evaluated as $A_{5-link}(t)=t^{-18/2}\text{Det}(S-tS^{T})=\left(\sqrt{t}-\frac1{\sqrt{t}}\right)^4$.

\section{Section II: Nodal line solutions of the Floquet system}
\subsection{General solutions}
 {We consider a general two-step quenching system described by}
\begin{eqnarray}
&& {H_1=h_1^0\mathbb{I}+\bm{h}_1\cdot\bm{\sigma},}\nonumber\\
&& {H_2=h_2^0\mathbb{I}+\bm{h}_2\cdot\bm{\sigma}}
\end{eqnarray}
 {with a overall Floquet period $T=T_1+T_2$, and the Hamiltonian of the system is $H=H_1$ for duration $T_1$ and $H = H_2$ for duration $T_2$.
Consider a specific time window from $t=T_1/2$ to $t=3T_1/2+T_2$, }
the Floquet operator takes the form of
\begin{eqnarray}
\hat{U}_T&=&e^{-iH_1T_1/2}e^{-iH_2T_2}e^{-iH_1T_1/2}\nonumber\\
&=&e^{i(h_1^0T_1+h_2^0T_2)}(r_0\mathbb{I}+i{\bm r}_U\cdot {\bm \sigma}),\label{evolution_op}
\end{eqnarray}
where
\begin{eqnarray}
r_0&=&\cos(|\bm{h}_1|T_1)\cos(|\bm{h}_2|T_2)\nonumber\\
&&-\sin(|\bm{h}_1|T_1)\sin(|\bm{h}_2|T_2)\bm{\hat{h}}_1\cdot\bm{\hat{h}}_2,\nonumber\\
{\bm r}_U&=&[\sin(|\bm{h}_1|T_1)\cos(|\bm{h}_2|T_2)\nonumber\\
&&+\cos(|\bm{h}_1|T_1)\sin(|\bm{h}_2|T_2)\bm{\hat{h}}_1\cdot\bm{\hat{h}}_2]\bm{\hat{h}}_1\nonumber\\
&&+\sin(|\bm{h}_2|T_2)[\bm{\hat{h}}_2-(\bm{\hat{h}}_1\cdot\bm{\hat{h}}_2)\bm{\hat{h}}_1],
\end{eqnarray}
with $\bm{\hat{h}}=\bm{h}/|\bm{h}|$. The unitarity of $\hat{U}_T$ leads to $|r_0|^2+|{\bm r}_U|^2=1$, hence one can always find a ${\bm r}_u$ satisfying $r_0=\cos{|{\bm r}_u|}$ and ${\bm r}_U=\sin{|{\bm r}_u|}$, with ${\bm \hat{r}}_u={\bm r}_u/|{\bm r}_u|$. Therefore the Floquet operator can be written as
\begin{eqnarray}
\hat{U}_T=e^{i(h_1^0T_1+h_2^0T_2)}(\cos{|{\bm r}_u|}\mathbb{I}+i{\bm \hat{r}}\sin{|{\bm r}_u|}\cdot {\bm \sigma}),\label{evolution_op}
\end{eqnarray}
and the effective Hamiltonian can be obtained as
\begin{eqnarray}
H_{\rm eff}&=&\frac{i}{T_1+T_2}\ln \hat{U}_T\nonumber\\
&=&-\frac{1}{T_1+T_2}[(h_1^0T_1+h_2^0T_2)\mathbb{I}-{\bm r}_u\cdot{\bm \sigma}].
\end{eqnarray}
Note that there is no cross product of Pauli matrices appearing in this procedure, so that if any sublattice symmetry is present in both $H_1$ and $H_2$, it will continue to present in the Floquet effective Hamiltonian. If we choose
\begin{eqnarray}
&&H_1=\bm{h}_1\cdot\bm{\sigma}=h_1^x\sigma_x+h_1^z\sigma_z,\nonumber\\
&&H_2=\bm{h}_2\cdot\bm{\sigma}=h_2^x\sigma_x+h_2^z\sigma_z,\nonumber\\
\end{eqnarray}
the resulting effective Hamiltonian shall still have no $\sigma_y$ term.  The two Floquet bands shall touch each other when the Pauli matrices disappear in Eq. (\ref{evolution_op}),  which yields the following band-touch conditions
\begin{eqnarray}
&&\bm{\hat{h}}_1=\bm{\hat{h}}_2,~|\bm{h}_1|T_1+|\bm{h}_2|T_2=n\pi,~\mathrm{or}\label{con01}\\
&&\bm{\hat{h}}_1=-\bm{\hat{h}}_2,~|\bm{h}_1|T_1-|\bm{h}_2|T_2=n\pi,\label{con02}
\end{eqnarray}
or the conditions
\begin{eqnarray}
T_1|{\bm{h}_1}|=n_1\pi,~~T_2|{\bm{h}_2}|=n_2\pi.\label{other_loops_sup}
\end{eqnarray}
Due to the periodicity of the quasienergies $\varepsilon$, the two Floquet bands may touch each other either at $\varepsilon=0$, which corresponds to an even $n$ or $n_1+n_2$; or at $\varepsilon=\pi/T$ with $T=T_1+T_2$, which corresponds to an odd $n$ or $n_1+n_2$. Finally, by taking $T_1=T_2=T/2$ and $\Delta_n=2n\pi/T$, we obtain
\begin{eqnarray}
&& {\bm{\hat{h}}_1=-\bm{\hat{h}}_2,~|\bm{h}_1|-|\bm{h}_2|=\frac{2n\pi}{T}=\Delta_n,}~\mathrm{or}\label{con11}\\
&& {\bm{\hat{h}}_1=\bm{\hat{h}}_2,~|\bm{h}_1|+|\bm{h}_2|=\frac{2n\pi}{T}=\Delta_n,}\label{con12}
\end{eqnarray}
from Eqs.~(\ref{con01}) and (\ref{con02}). These two conditions are Eq.~(2) and Eq.~(3) in the main text and are of our main interest.
Eqs. (\ref{other_loops_sup}) also give some nodal loops, however our numerical results show that they do not generate any new linkage.

\subsection{Simplified expressions for nodal solutions}
In this section we derive the expression of $g_n({\bm k})$ used in the main text and also compute the linkage of loops given by $g_n({\bm k})=0$ with different $n$. First we define a series of complex functions $f_{n}({\bm k})=z^2({\bm k})-w^2({\bm k})-\Delta_n$ with
\begin{eqnarray}
z(\bm{k})=\alpha_x+i\beta_z,~~w(\bm{k})=\beta_x+i \alpha_z,
\end{eqnarray}
\begin{eqnarray}
\alpha_i=\frac{h_1^i}{\sqrt{{|\bm{h}_1|}}},~~
\beta_i=\frac{h_2^i}{\sqrt{{|\bm{h}_2|}}}.
\end{eqnarray}
Requiring $f({\bm k})=0$ yields\begin{eqnarray}
{\rm Re}[f_{n}({\bm k})]=|{\bm h}_1|-|{\bm h}_2|-\Delta_n=0,\label{fk_R}\\
{\rm Im}[f_{n}({\bm k})]=2(h_1^xh_2^z-h_1^zh_2^x)/\sqrt{|{\bm h}_1||{\bm h}_2|}=0,\label{fk_I}
\end{eqnarray}
we can see that Eq. (\ref{fk_R}) recovers the second part of Eq. (\ref{con11}), and Eq. (\ref{fk_I}) is equivalent to $\bm{\hat{h}}_1=\pm\bm{\hat{h}}_2$.

Next we rewrite the complex functions as
\begin{eqnarray}
f_{n}({\bm k})=g^+_n({\bm k})g^-_n({\bm k})
\end{eqnarray}
with
\begin{eqnarray}
g^{\pm}_n({\bm k})&=&z({\bm k})\pm w'_{n}({\bm k}),
\end{eqnarray}
and $w'_{n}({\bm k})=\sqrt{w^2({\bm k})+\Delta_n}$. Since $\Delta_n$ is a real number, $w'_n({\bm k})$ and $w({\bm k})$ are always in the same quadrant in the complex plane, i.e.
\begin{eqnarray}
{\rm sgn}[{\rm Re}(w'_n({\bm k}))]={\rm sgn}[{\rm Re}(w({\bm k}))],~~~~
{\rm sgn}[{\rm Im}(w'_n({\bm k}))]={\rm sgn}[{\rm Im}(w({\bm k}))].
\end{eqnarray}
Therefore $g^{\pm}_n({\bm k})=0$ can only satisfy $\bm{\hat{h}}_1=\mp\bm{\hat{h}}_2$ respectively, and Eqs.~(\ref{con11}) is recovered by $g^+_n({\bm k})=0$. We hence arrive at $g_n({\bm k})=0$ to express Eqs.~(\ref{con11}), with
\begin{eqnarray}
g_n({\bm k})\equiv g^{+}_n({\bm k})=z({\bm k})+ \sqrt{w^2({\bm k})+\Delta_n}.\label{gdelta_sup}
\end{eqnarray}
Thus, we have first used the compact condition, i.e., $f_n({\bm k})=0$ and then reduce it to $g_n({\bm k})=0$ to express our nodal solutions yielded from Eqs.~(\ref{con11}). 

\subsection{Pairwise linkage between nodal loops given by $g_n({\bm k})=0$}
\begin{figure}
\includegraphics[width=0.7\linewidth]{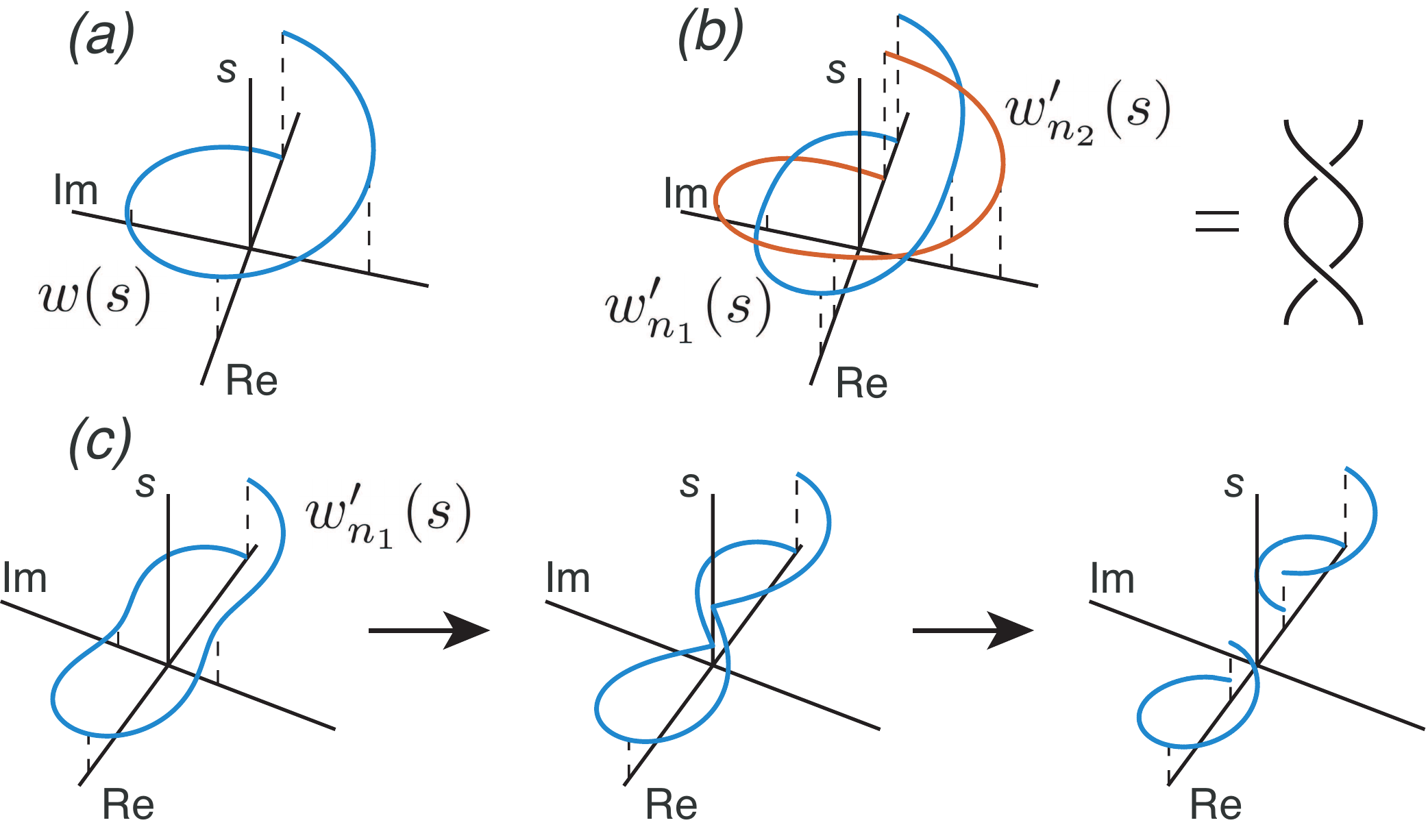}
\caption{(a) Schematic of how a loop defined by  $z(\bm{k})=0$ is linked with $w(\bm{k})=0$, as visualized upon a mapping that
maps the former to the vertical axis. (b) Under the same mapping, loops defined by $\sqrt{w^2(\bm{k})+\Delta_{n_1}}=0$ and $\sqrt{w^2(\bm{k})+\Delta_{n_2}}=0$ with small $\Delta_{n_2}$
and small $\Delta_{n_1}$ are found to be always linked. (c)  The transition of $w_{n_1}'(s)$ when increasing $\Delta_{n_1}.$}
\label{figS1}
\end{figure}

In order to unveil the linkage of loops given by $g_n({\bm k})=0$ with different $n$, we first define a linking number as
\begin{eqnarray}
\nu_{h_1,h_2}=\frac{1}{2\pi^2}\int_{\rm{BZ}}d^3 \bm{k}\,\epsilon_{abcd}N_{a}\partial_{k_x} N_{b} \partial_{k_y} N_{c} \partial_{k_z} N_{d},\label{linking}
\end{eqnarray}
with $N_i$ the $i$-th component of the vector $(h_1^x,h_1^z,h_2^x,h_2^z)$ normalized to unit length \cite{Ezawa_Hopf,Yan_Hopf}. This linking number reflects the linkage of the original nodal loops of the static Hamiltonian $H_1$ and $H_2$. In this study, we assume $\nu_{h_1,h_2}=1$, where the static nodal loops are linked, as illustrated in Fig.~1(b) in the main text.  Because $z(\bm{k})$ and $w(\bm{k})$ are in essence rescaled and rearranged ${\bm h}_1$ and ${\bm h}_2$, we have that for the parameter regime associated with Fig.~1(b),  the two loops defined by $z(\bm{k})=0$ and $w(\bm{k})=0$ are linked. For a better visualization of such a linkage between $z(\bm{k})=0$ and $w(\bm{k})=0$, we introduce an angular coordinate $s\in[0,2\pi]$ along the loop $z(\bm{k})=0$ and then map $z(\bm{k})=0$ to the vertical axis.  Under the same mapping $w(\bm{k})=0$ becomes a curve $w(s)$ that winds around the vertical axis. In Fig.~\ref{figS1}(a) we show a sketch of $w(s)$ in the angular coordinate system.

Clearly, within the small-$|\Delta_n|$ regime, $g_{n=n_1}(\bm{k})=0$ and $g_{n=n_2}(\bm{k})=0$ yield two different nodal loops for $n_1\neq n_2$. These two loops can be regarded as deformations of $g_{0}(\bm{k})=0$ if $\Delta_{n_1}$ and $\Delta_{n_2}$ are small enough. Under the same mapping illustrated in Fig.~\ref{figS1}(a), $\sqrt{w^2(\bm{k})+\Delta_{n_1}}=0$ and $\sqrt{w^2(\bm{k})+\Delta_{n_2}}=0$ are respectively mapped to $w_{n_1}'(s)$ and $w_{n_2}'(s)$, as shown in Fig.~\ref{figS1}(b). It is found that $w_{n_1}'(s)$ and $w_{n_2}'(s)$ are linked with unity linking number. One can then conclude that the nodal loops $g_{n_1}(\bm{k})=0$ and $g_{n_2}(\bm{k})=0$ are always pairwise linked for sufficiently small $\Delta_n$. This result is also confirmed by our numerical calculation of the linking number of these two loops, defined as the same as in Eq.~({\ref{linking}}), but with $N_a$ the $a$-th component of the vector $(\Re[g_{n_1}],\Im[g_{n_2}],\Re[g_{n_2}],\Im[g_{n_2}])$ normalized to unit length.

\subsection{Transitions of the nodal solutions}

One wonders for a given finite driving period $T$, how many pair-wise linked loops as described above
can be generated from the condition $g_{n}(\bm{k})=0 $ with a varying $n$.
This question is equivalent to ask at what condition the linkage of the loops will can be changed.
To answer it, first we need to locate the boundary of the above-mentioned small-$|\Delta_n|$ regime.
From Eq.~(\ref{gdelta_sup}), it is seen that if $n>0$ keeps increasing, the points of $w'_n(s)$ originally on the imaginary axis moves to origin when $\Delta_n=(\Im [w(s)])^2$, and then splits into
 two on the real axis moving at opposite directions when $\Delta_n$ keeps increase, as depicted in Fig. \ref{figS1}(c). Thus, beyond the boundary $\Delta_n=(\Im [w(s)])^2$, a nodal loop $g_{n}(\bm{k})=0$ as a smooth deformation from $g_{n=0}(\bm{k})=0$ starts to break into disjoint parts, and the end points of these parts satisfy $\Re[w(s)]=0$.
Specifically, in our model this situation occurs when $\Re[w({\bm k})]=\beta_x=0$ and $\Delta_{n}>(\Im[w({\bm k})])^2=\alpha_z^2$, under which $g_{n}({\bm k})=0$ requires $\alpha_x=\pm\sqrt{\Delta_{n}-\alpha_z^2}$ and $\beta_z=0$. In other words, the nodal loop of $g_{n}({\bm k})=0$ splits into several segments, and the end points of each segment is given by
\begin{eqnarray}
{\bm h}_2({\bm k})=0,~~|{\bm h}_1({\bm k})|=\Delta_{n}.\label{joining_+n}
\end{eqnarray}
A similar analysis applies to cases with $n<0$, while the condition for the end points is given by
\begin{eqnarray}
{\bm h}_1({\bm k})=0,~~|{\bm h}_2({\bm k})|=-\Delta_{n}.\label{joining_-n}
\end{eqnarray}
These two conditions satisfy the band-touching conditions Eqs.~(\ref{con11}) and (\ref{con12}) simultaneously, with $n=|n|$ for Eq. (\ref{con12}).
The overall picture is the following: the band-touching condition $g_{\pm n}({\bm k})=0$ with $|\Delta_n|$ beyond a critical value $\Delta_c$ will give several disjoint nodal lines, and the end points of these lines satisfy both conditions of Eq.~(\ref{con11}) and Eq.~(\ref{con12}). In this regime, due to the continuity of the Hamiltonian, Eq.~(\ref{con11}) shall also give some nodal lines that connect these end points, and form closed nodal loops together with the solution of Eq.~(\ref{con12}). Such an example is depicted by Fig.~3(a)-(c) in the main text.

This transition does not change the linkage of the loops, as it does not involve touching of the overall loops composed by the solutions of Eq.~(\ref{con11}) and Eq.~(\ref{con12}). However, the emergence of solutions of Eq.~(\ref{con12}) make further topological transition of the loops possible in our system.
In the small-$|\Delta_n|$ regime, the nodal loops are given by $g_{n}({\bm k})=0$, and two such conditions with different $n$ cannot be satisfied at the same time. Therefore, the linked loops always remain linked in this regime. However, from above analysis we can see that when $|\Delta_n|$ increases, the linked nodal loops are partially (or completely) given by the solutions of Eq.~(\ref{con12}) with the same $n=|n|$, and there is not any restriction forbidding them from touching.
In our system, we observe that the two linked loops for a given $n$ of Eq. (\ref{con12}) touch at ${\bm k}_0=(0,0,0)$ when $T$ decrease to $\pi|n|/(\cos{\phi}+1- {\mu})$, and unlink with smaller $T$.
Here we give a detailed analysis of this transition.

From above we learn that as the nodal loop solutions to Eq.~(\ref{con11}) split, there exist end points along this solution that also satisfy Eq.~(\ref{con12}). As such  we define $g'_{n}({\bm k})=0$ as solutions to Eq. (\ref{con12}) which connect and eventually replace the nodal loops given by $g_{n}({\bm k})=0$.
Likewise,  we define $g'_{-n}({\bm k})=0$ as solutions to Eq.~(\ref{con12}) which connect and eventually replace the nodal loops given by $g_{-n}({\bm k})=0$.
To better understand $g'_{\pm n}({\bm k})=0$
we  rewrite Eq. (3) of the main text as
\begin{eqnarray}
{\bm h}_2({\bm k})=c_+({\bm k}){\bm h}_1({\bm k}),~~[1+c_+({\bm k})]|{\bm h}_1({\bm k})|=\Delta_n,\label{con_+}
\end{eqnarray}
when considering $g'_{ n}({\bm k})=0$ and
\begin{eqnarray}
{\bm h}_1({\bm k})=c_-({\bm k}){\bm h}_2({\bm k}),~~[1+c_-({\bm k})]|{\bm h}_2({\bm k})|=\Delta_n.\label{con_-}
\end{eqnarray}
when considering $g'_{-n}({\bm k})=0$.  With this convention, it is evident that
\begin{eqnarray}
c_-({\bm k})=\frac{1}{c_+({\bm k})},
\end{eqnarray}
because both expressions above refer to the same condition for nodal lines.  Further, according to our discussions above, we have that (i) $c_+({\bm k})$ is a scalar non-negative function of ${\bm k}$, which is zero at the end points of $g'_{ n}({\bm k})=0$; and (ii) $c_{-}({\bm k})$ is a scalar non-negative function of ${\bm k}$, which is zero at the end points of $g'_{-n}({\bm k})=0$.

It is now more convenient to examine when solutions given by $g'_{ n}({\bm k})=0$ and $g'_{-n}({\bm k})=0$ may touch. Consider their respective solutions
\begin{eqnarray}
&&\{{\bm k}_+:~g'_{n}({\bm k}_+)=0\}\\
&&\{{\bm k}_-:~g'_{-n}({\bm k}_-)=0\}.
\end{eqnarray}
The task is to analyze if the above two sets of solutions have any overlap.  Suppose
$\max[c_+({\bm k}_+)]\max[c_-({\bm k}_-)]<1$.  Because $c_+({\bm k}_+)\leqslant\max[c_+({\bm k}_+)]$, then for any ${\bm k}_+$,
\begin{eqnarray}
c_-({\bm k}_+)=\frac{1}{c_+({\bm k}_+)}\geqslant \frac{1}{\max[c_+({\bm k}_+)]}>\max[c_-({\bm k}_-)],
\end{eqnarray}
which does not allow any overlap between ${\bm k}_+$ and ${\bm k}_-$.  Thus, the two solutions from $g'_{ n}({\bm k})=0$ and $g'_{-n}({\bm k})=0$ cannot touch each other in the regime of $\max[c_+({\bm k}_+)]\max[c_-({\bm k}_-)]<1$.

At the point when the nodal loop solutions to Eq.~(\ref{con11}) begin to split, Eq.~(\ref{con12}) only has solutions at the splitting points, where $\max[c_+({\bm k}_+)]\max[c_-({\bm k}_-)]=0$.  As we tune up $\max[c_+({\bm k}_+)]\max[c_-({\bm k}_-)]$ by increasing $\Delta_n$, there is a possibility to reach $\max[c_+({\bm k}_+)]\max[c_-({\bm k}_-)]=1$.   Using the similar analysis as above, then it becomes possible for the two solutions given by $g'_{ n}({\bm k})=0$ and $g'_{-n}({\bm k})=0$ to touch.
 In our model, the Hamiltonian satisfies ${\bm h}_1(k_x,k_y,k_z)={\bm h}_2(-k_x,k_z,k_y)$. Under this symmetric relation, Eq.~(\ref{con_-}) can be transformed to
\begin{eqnarray}
{\bm h}_2(\mathcal{R}{\bm k})=c_{-}(\mathcal{R}{\bm k}){\bm h}_1({\mathcal{R}\bm k}),~~[1+c_-(\mathcal{R}{\bm k})]|{\bm h}_1(\mathcal{R}{\bm k})|=\Delta_n
\end{eqnarray}
with $\mathcal{R}(k_x,k_y,k_z)=(-k_x,k_z,k_y)$.
In other words, for any ${\bm k}_+$ with a specific value of $c_+({\bm k}_+)$, there must be a ${\bm k}_-=\mathcal{R}{\bm k}_+$ with $c_-(\mathcal{R}{\bm k})=c_+({\bm k})$. Therefore the condition $\max[c_+({\bm k}_+)]\max[c_-({\bm k}_-)]=1$ becomes $\max[c_+({\bm k}_+)]=\max[c_-({\bm k}_-)]=1$, and the touching point ${\bm k}_0$ shall satisfy $c_{\pm}({\bm k}_0)=1$, i.e. ${\bm h}_1({\bm k}_0)={\bm h}_2({\bm k}_0)$.

\begin{figure}
\includegraphics[width=0.7\linewidth]{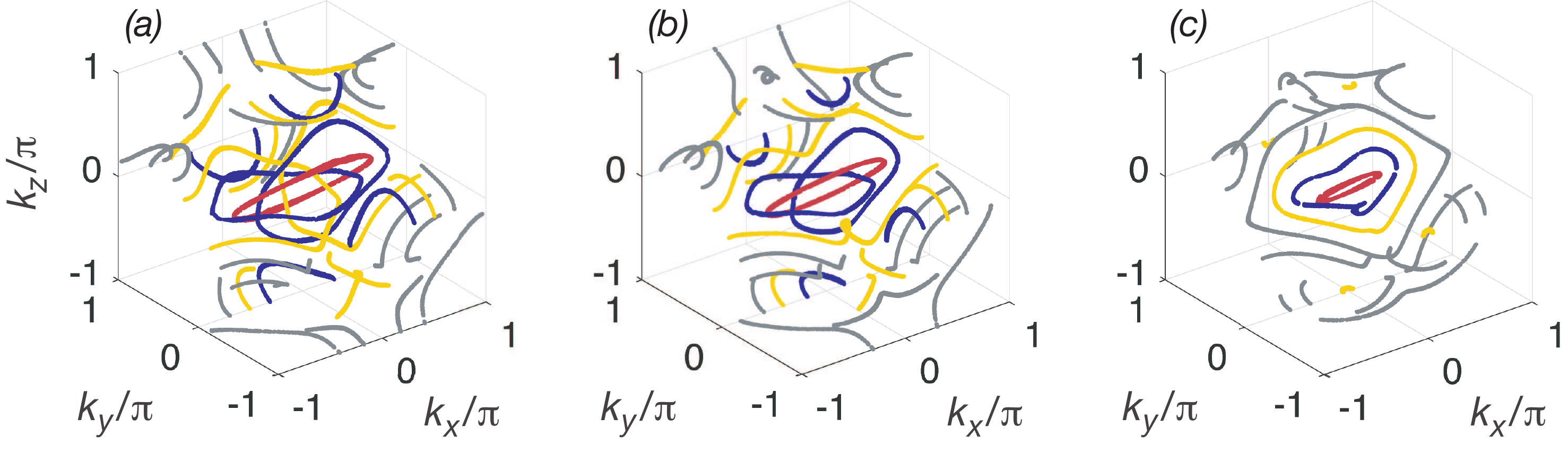}
\includegraphics[width=0.7\linewidth]{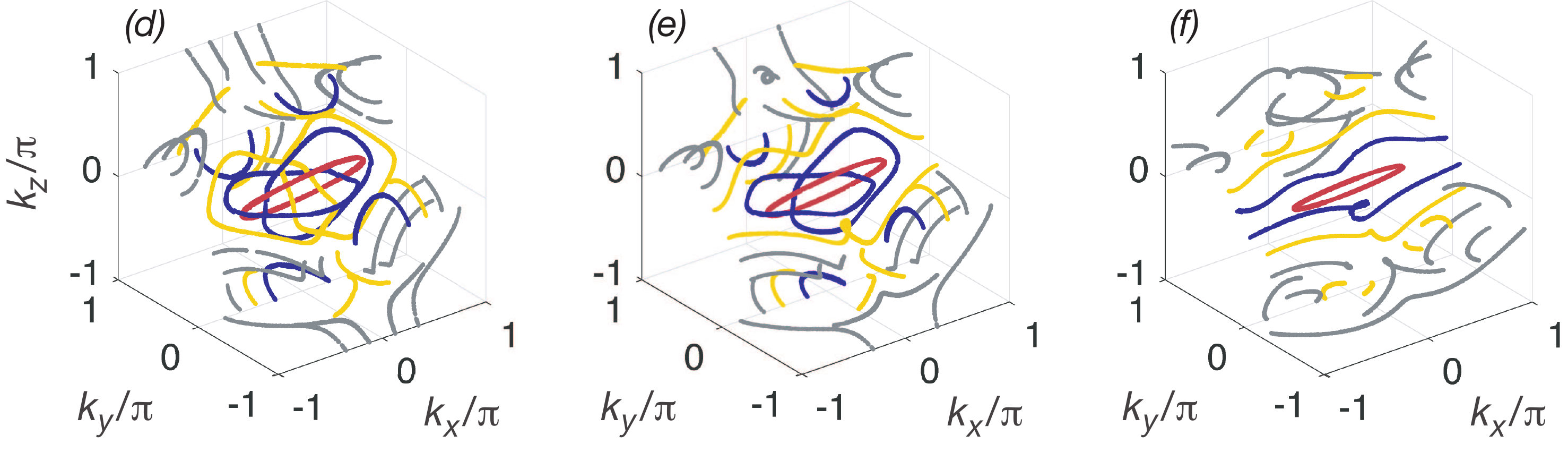}
\caption{Nodal lines at $\varepsilon=0$ given by Eqs. (\ref{con11}) and (\ref{con12}) for the same system as discussed in the main text. The upper panels are for $T=5\pi$, $\phi=\pi/4$, and (a) $ {\mu}=0.8$; (b) $ {\mu}=1$ and (c) $ {\mu}=1.5$. The lower panels are for $T=5\pi$, $ {\mu}=1$, and (d) $\phi=\pi/6$; (e) $\phi=\pi/4$ and (f) $\phi=5\pi/12$. The red, blue and yellow color are for $n=0$, $n=\pm2$ and $n=\pm4$ respectively, and gray lines represent nodal lines obtained for larger even $n$ (up to $8$).}\label{FigS2}
\end{figure}

As a conclusion, the critical value of $|\Delta_n|=\Delta_c$ shall be given by
\begin{eqnarray}
\Delta_{c}={\rm min}(2|{\bm h}_1({\bm k})|)
\end{eqnarray}
when ${\bm h}_1({\bm k})={\bm h}_2({\bm k})$ is satisfied.
Interestingly, the above condition applied to our model yields a single touching point at ${\bm k}_0=(0,0,0)$ with $\Delta_c=2(\cos{\phi}+1- {\mu})$, with the critical value of $T$ found to be
\begin{eqnarray}
T=\frac{2\pi|n|}{\Delta_c}=\frac{\pi|n|}{(\cos{\phi}+1- {\mu})}.
\end{eqnarray}
 Finally, as the critical value also depends on $m$ and $\phi$, we also present the transitions induced by changing $\mu$ and $\phi$ in Fig.~(\ref{FigS2}) as a side result, complementing the results in the main text focusing on transitions induced by a varying $T$.

\subsection{Winding number characterizations of the nodal line touching points}
\begin{figure}
\includegraphics[width=0.7\linewidth]{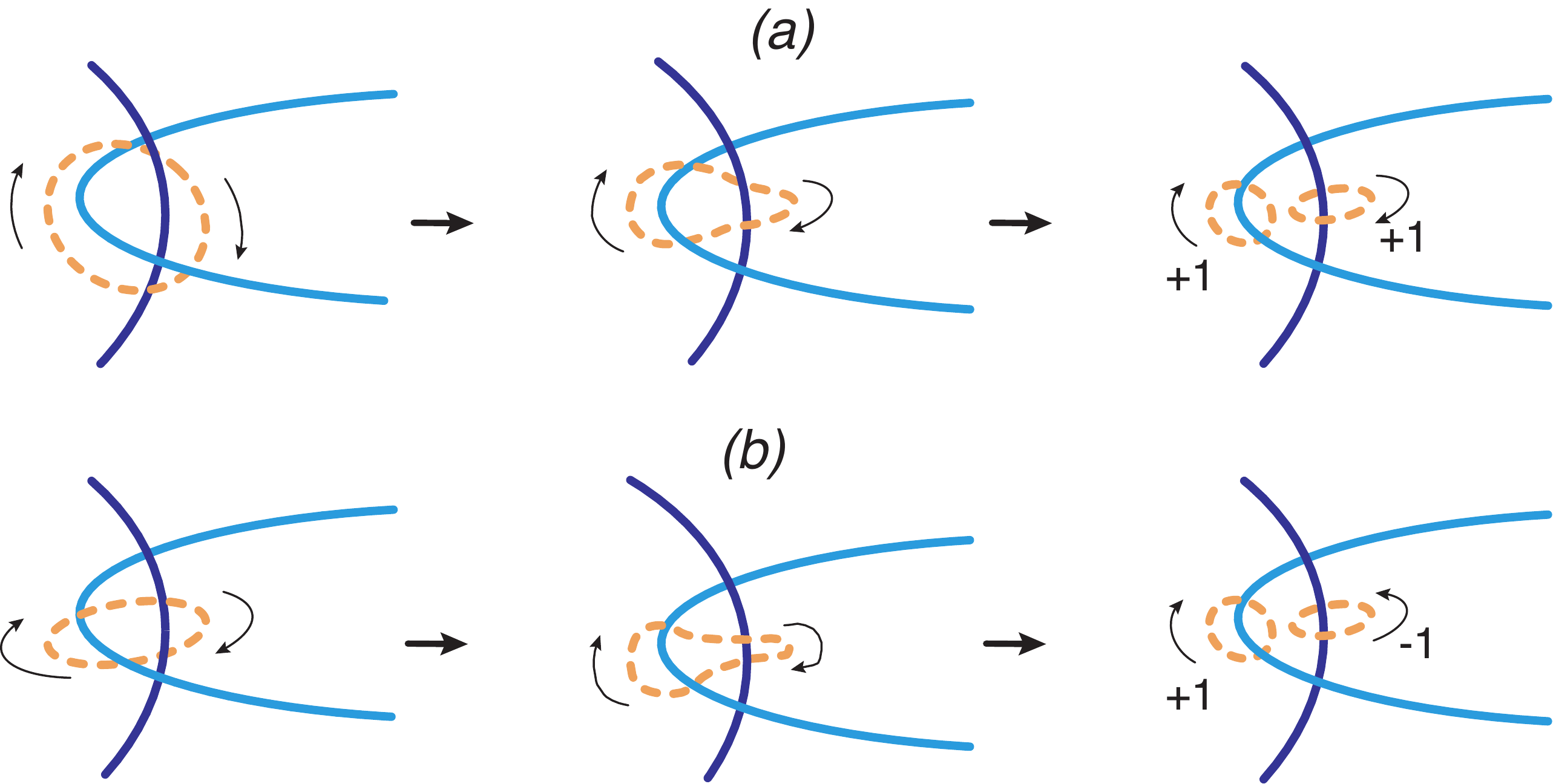}
\caption{Sketch of a trajectory with respect to a pairwise link in two different directions. The light and dark blue lines indicate the two nodal loops of a pairwise link, and the yellow dash lines indicate the trajectories along which the winding number is calculated. Both trajectories in (a) and (b) can be deformed into two small circles enclosing a part of each loop of a pairwise link, with the case (b) featuring two small closed trajectories having opposite winding directions.}\label{figS3}
\end{figure}

In a NLSM, a nodal loop can be characterized by a winding number of $\nu=\pm1$, which indicate a $\pi$ Berry phase, along a trajectory linked with the loop. The plus or minus sign of the winding number assigns a winding direction to the nodal loop.
For a pairwise link, one may consider a trajectory enclosing the link, and its winding number $\nu_{\rm pair}$ is contributed by the winding numbers of both loops, as the trajectory can be continuously deformed into two linked with the two loops respectively. The value of $\nu_{\rm pair}$ may take either summation or the difference of the single-loop winding numbers, depending on whether the two trajectories are connected in the same or opposite directions in the deformation, as shown in Fig. \ref{figS3}.

From this aspect, for a pairwise link where the winding directions of the two loops are unknown, we need to consider two trajectories enclosing these loops with different directions, one of them shall give a winding number of $\pm2$ while the other gives zero. However, to determine such trajectories, we first need to locate the position of the loops, which varies greatly when tuning the parameters in our system. Nevertheless, for each pair of loops with a given $|n|$, their linkage changes at a different $T=\pi|n|/(\cos{\phi}+1- {\mu})$, while the touching of the two loops always occurs at ${\bm k}_0=(0,0,0)$.
Therefore we consider two small trajectories
\begin{eqnarray}
&&c_1:~k_x=r\sin{\theta},~k_y=k_z=r\cos{\theta}\nonumber\\
&&c_2:~k_x=r\sin{\theta},~k_y=-k_z=r\cos{\theta},\label{winding_path}
\end{eqnarray}
which enclose ${\bm k}_0$ in two planes perpendicular to each other. Here $r$ is the radius of the trajectories, and $\theta$ is the phase angle. For a given value of $r$, the pairwise link given by $|n|$ shall falls within both trajectories when $T$ is in a small regime around $T=\pi|n|/(\cos{\phi}+1- {\mu})$. Therefore in this regime one of the two corresponding winding numbers $\nu_1$ and $\nu_2$ shall takes the value of $\pm2$, while the other is zero. In addition, when $r$ tends to zero, this regime shall shrink into a single point at $T=\pi|n|/(\cos{\phi}+1- {\mu})$. Hence the jump of winding number $\nu_1$ and $\nu_2$ shows a signal of the transition where the number of linked nodal loops change by two, as shown in Fig. (4) in the main text.  Interestingly, we find that for the pairwise link at quasi-energy $\varepsilon=0$ ($\pi/T$) with odd (even) $n$, the transition is reflected by the jump of $\nu_{1(2)}$, and $\nu_{2(1)}$ is always zero. Moreover, we have $\nu_1=(-1)^{(n-1)/2}$  for odd $n$, and $\nu_2=(-1)^{(n-2)/2}$ for even $n$. These results indicate that the loops given by different $n$  have different winding directions, even though they
have superficially the same shape and move in the same pattern when tuning $T$.

 {In experimental measurements of the Berry phase related winding number,  a winding number of 2 corresponds to a zero Berry phase. However, we can always divide the trajectory where we calculate the winding number into two smaller ones, each yielding a winding number of 1, corresponding to a Berry phase of $\pi$. Then we can detect the jumps of winding number in Fig. 4 of the main text by the jumps of the two Berry phase from 0 to $\pi$.}

Finally, we note that for even value of $n$, when $T\lesssim\pi|n|/(\cos{\phi}+1- {\mu})$ where the $|n|$-th pairwise link is broken into two separated loops,
there is another nodal loop given by Eq. (\ref{other_loops_sup}), instead of Eqs.~(\ref{con11}) or (\ref{con12}). This nodal loop is near the ${\bm k}_0$ point and connects the two separated loops. When decreasing $T$, the jump of winding number $\nu_2$ by $2$ occurs when this nodal loop crosses the trajectory $c_2$. On the other hand, when increasing $T$, this loop shrinks into a point and disappear at $T=\pi|n|/(\cos{\phi}+1- {\mu})$, leaving only the pairwise link when $T>\pi|n|/(\cos{\phi}+1- {\mu})$. Therefore the overall linkage is not affected by this extra loop, and the emergence or breaking of $|n|$-th pairwise link with even $n$ can still be well characterized by the jump of $\nu_2$.

 {
\section{Section III: Experimental characterization of linked nodal loops}}

 {Possible experimental measurements of linked nodal loops are of great interest by themselves.  The recent high profile characterizations of nodal lines were based on electronic materials, where the Fermi surface and hence nodal structure can be mapped via angle-resolved photoemission spectroscopy (ARPES) measurements. Some of these nodal materials include $\text{PbTaSe}_{2}$~\cite{bian2016topological}, $\text{TiB}_{2}$~\cite{liu2017experimental} and $\text{ZrB}_{2}$~\cite{lou2018experimental}. Very recently, nodal chains were also detected in a specially designed photonic system via analogous angle-resolved transmission measurements~\cite{yan2018experimental}.}

 {
By contrast, our proposed platform is tailored for cold-atom experiments, for which the detailed band structure and even pseudospin texture can be more easily mapped out in momentum space. Given that our nodal loop linkages are significantly more intricate than those proposed in any other realistic system, it is difficult for any single observable to fully characterize their topological structure. Below we outline a few distinct but complementary experimental approaches that together provide a fingerprint for a given set of linked nodal loops for our platform.}

\bigskip

{ {\subsection{A. Tomography of Berry phase, nodal structure and pseudospin}}

 {The detailed loop structures of our nodal lines are most directly probed via Berry phase measurements along suitably designed trajectories (see also the main text). Indeed, such measurements along arbitrary trajectories in momentum (reciprocal) space trajectories have been performed in a few recent landmark experiments using cold-atom systems.}

 {One experiment (Ref.~\cite{li2016bloch}) measured the Berry phase by accelerating cold atoms along arbitrary desired trajectories (Wilson lines) in reciprocal space. The shape of the Wilson line can be controlled by systematically varying two of the beams defining the optical lattice. Since the transport along the Wilson line is adiabatic compared to our quenching rate, a similar approach can be employed to measure the Berry phases associated with our nodal linkages.}

 {Almost contemporaneously, the Berry phase in a Floquet cold-atom system was also measured via a time-of-flight approach~\cite{flaschner2016experimental}. By appropriately switching off the optical lattice after the desired state is prepared, the atoms are released in a manner such that the momentum-resolved distributions of their pseudospin components can be independently measured~\cite{alba2011seeing,hauke2014tomography}. This information allows mapping (tomography) of the Berry curvature and even the detailed shape of the Fermi surface~\cite{kohl2005fermionic}. This approach is also applicable to our our setup which consists of qualitatively similar driven cold atoms, thereby allowing for a detailed map of its exotic nodal structure.   Most relevant to this work, we highlight that there is  already a powerful tool available to measure Berry phase related winding numbers in both static and periodically driven cold-atom systems.}


\bigskip

 {\subsection{B. Nodal structure characterization through non-linear response}}

 {Another experimental signature of nodal loop structures  is their characteristic non-monotonic response to external fields, which also results in characteristic frequency multiplication behavior when the external field is fluctuating. As explained in Ref.~\onlinecite{lee2015negative}, the peculiar topology of the distribution of the occupied states around a nodal loop results in a non-monotonic semi-classical response within the plane of the loop. When multiple loops are present, we expect to measure non-linear responses or their associated frequency multiplying effects in various relevant directions. This approach will provide useful information on the configuration of the nodal loops independent of the abovementioned tomography data, even if it alone cannot discern the exact positions of the loops. Indeed, similar ideas have been employed for studying other arguably simpler nodal systems like Dirac/Weyl systems~\cite{morimoto2016semiclassical}.}

 {For our system with electrically neutral cold atoms, the external fields may be engineered with laser-atom interactions. It is also by now well-established that the response in such cold-atom systems can be measured by tracking the center-of-mass position of a cloud of atoms~\cite{price2016measurement}. Indeed, this was demonstrated as a topological charge pump in a recent high profile experiment~\cite{lohse2018exploring}. Such measurements are expected to generalize to 3 dimensions, as will be relevant for our nodal loops, with realistic proposals both from the same group~\cite{petrides20186d} and from one of our authors~\cite{lee2018electromagnetic}.}

\bigskip

 {\subsection{C. Possible reconstruction of nodal loop linkage from momentum-resolved surface measurements}}

 {As emphasized in our work, a key physical consequence of the nontrivial topology of the nodal structures is the appearance of protected degenerate surface ``drumhead'' states shaped according to the surface projections of the bulk Seifert surface. By combining information about the drumhead states on surfaces oriented along different directions, one can unambiguously reconstruct the 3D Seifert surface within the momentum-space bulk, and hence uniquely reveal the geometry and topology of the nodal loop linkages.  Note that while the nodal loops do not uniquely define the Seifert surface, the latter uniquely determines the nodal loop configuration.}

 {To reconstruct the linkage of the nodal loops, one possible route is based on momentum-resolved surface measurements. As a very new research direction,  drumhead state measurements have already been demonstrated in a photonics setting~\cite{yan2018experimental}. Furthermore, quasiparticle interference measurements appear highly promising for mapping out the drumhead states~\cite{biderang2018drumhead,PhysRevB.95.140202}. This technique was already employed to map the Fermi arcs of nodal points~\cite{inoue2016quasiparticle}.  Because such measurements are based on the momentum space auto-correlation function~\cite{kourtis2016universal}, they are also well-suited for probing the surface DOS profile in momentum space (see also subsection D.)}

\begin{figure}
\includegraphics[width=1\linewidth]{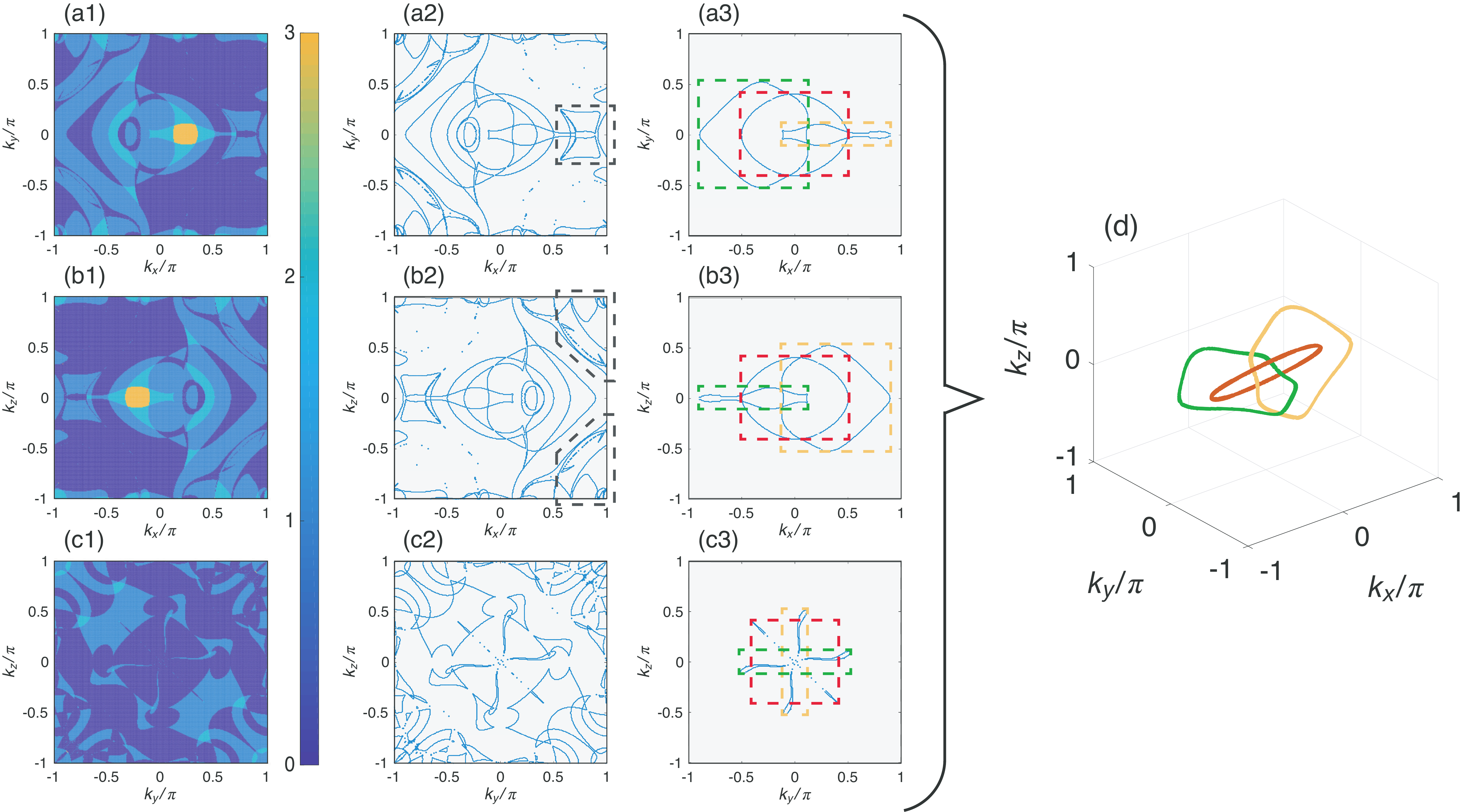}
\caption{Extracting the linkage of nodal loops from the profiles of ``drumhead" surface states. (a1)-(c1) depict the number of pairs of surface states with OBC along $z$, $y$, and $x$ directions respectively. (a2)-(c2) present the boundaries between the regions with different number of surface states. These boundaries are determined by the projections of the nodal loops.  The loops within the dark dashed lines in (a2) are actually separated from the ones near the center along $k_z$ direction, as shown in (b2). (a3)-(c3) depict the three possibly linked loops extracted from (a2)-(c2), with each color representing one nodal loop. (d) the linked nodal loops reconstructed from (a3)-(c3). The parameters are $\phi=\pi/4$, $\mu=1$, and $T=3\pi$, and we consider a system with $300$ unit cells (i.e. $600$ lattice sites) along the direction with OBC.
 }\label{fig_SS}
\end{figure}

 {Assuming that the number of pairs of drumhead states can be resolved in momentum, it becomes possible to reconstruct the topological linkage between different nodal loops.  In Fig.~\ref{fig_SS} we offer an example to illustrate how to extract the nodal loops and their linkage relations from the number of pairs of the ``drumhead" surface states. Here we take $T=3\pi$ and focus on the surface states at zero quasi-energy. Fig. \ref{fig_SS}(a1) to (c1) show the numerical results of surface states with open boundary condition (OBC) along $z$, $y$, and $x$ directions respectively. The different colors indicates the number of pairs of near-zero modes on the concerned surfaces under given momentum values.  Next, we read out the boundaries of these surface states, as shown in \ref{fig_SS}(a2) to (c2).  These boundaries are determined by the projections of the nodal loops onto the concerned surfaces along particular directions.  It can be seen that there may be some complex linkage near the central region of the Brillouin zone. The next task is to distinguish the linked loops from the unlinked ones. To do so one needs to patiently combine the information from all the three cases. For example, the loops within the dark dashed lines in (a2) seem to be linked to the loops near the center. However, consider the same regime of $k_x$ in (b2), the loops are all separated from the central ones. Therefore we can conclude that those loops in the dark dashed square in (a2) do not contribute to the total linkage of our system. After analyzing all the projections/boundaries one by one, even with just brute-force one can eventually obtain the relative positions of the three nodal loops near the center, as shown in Fig. \ref{fig_SS}(a3) to (c3). Finally, by matching the projections of these loops obtained along different directions, we can successfully reconstruct their linkage relation in the 3D Brillouin zone, as shown in Fig. \ref{fig_SS}(d).}

\

 {\subsection{D. Density of states and its conductance signature}}

\begin{figure}
\includegraphics[width=0.9\linewidth]{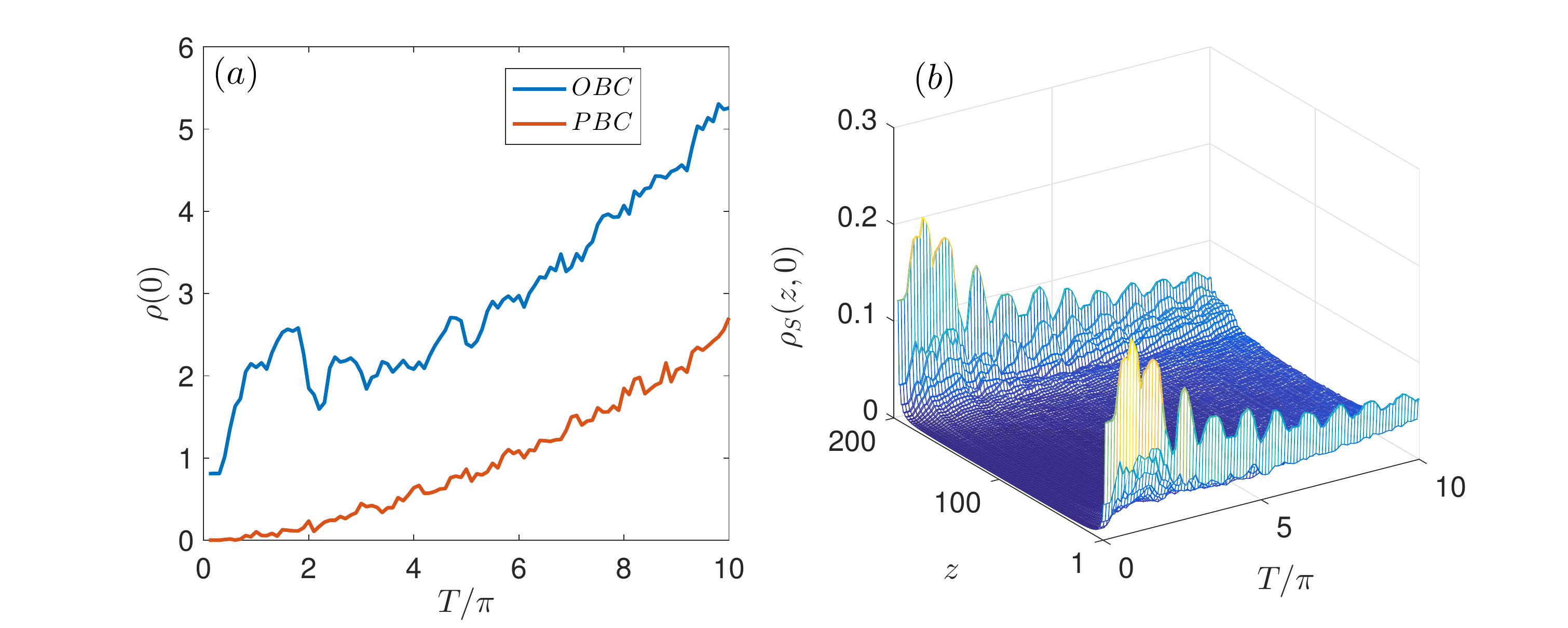}
\caption{The DOS and SDOS at zero energy as a function of the time-period $T$. We choose $\sigma=0.001$ in the numerical calculation.
The system's size is $N_x=50$, $N_y=50$, $N_z=200$. Other parameters as $\mu=1$ and $\phi=\pi/4$.}\label{fig_DOS}
\end{figure}

 {In addition to momentum-resolved surface measurements that can provide us more detailed information about the nodal loops, the collective DOS $\rho(\varepsilon)$ at quasi-energy $\varepsilon=0$ or $\pi/T$ can also serve as a signal of the nodal loops and ``drumhead" surface states. As an example, here we consider OBC along $z$ direction, and $\rho(\varepsilon_0)$ is defined as
\begin{eqnarray}
\rho(\varepsilon_0)=\frac{1}{N_xN_yN_z}\sum_{k_x,k_y,l} \delta(\varepsilon_0-\varepsilon_{k_x,k_y,l}),
\end{eqnarray}
where $N_a$ is the number of lattice sites in $a$ direction, and  $\varepsilon_{k_x,k_y,l}$ is the quasi-energy of the $l$-th state at given $k_x$ and $k_y$.
Because $\rho(\varepsilon_0)$ captures contributions from both the surface and the bulk, we also consider a surface DOS (SDOS) as follows:
\begin{eqnarray}
\rho_S(z,\varepsilon_0)=\frac{1}{N_xN_yN_z}\sum_{k_x,k_y,l} \delta(\varepsilon_0-\varepsilon_{k_x,k_y,l})\psi^*_{k_x,k_y,l}(z)\psi_{k_x,k_y,l}(z)
\end{eqnarray}
which provides spatial-resolved information along $z$ direction. Here $\psi_{k_x,k_y,l}(z)$ is defined as the amplitude of the wave-function of the $l$-th state at given $k_x$ and $k_y$ at the $z$-th lattice site.
}

 {Numerically, we approximate $\delta(x)$ by a Gaussian function $\frac{1}{\sqrt{\pi \sigma^2}}{\rm exp}(-\frac{x^2}{\sigma^2})$, which approaches the $\delta$-function when $\sigma\rightarrow0$.  For a finite-size system, the nodal loops and surface states are not exactly at zero (or $\pi/T$) quasi-energies,  and as such the chosen value of $\sigma$ cannot be too large or too small.}

 {In Fig. \ref{fig_DOS}, we show the DOS $\rho(0)$ and SDOS $\rho_S(z,0)$ as a function of the time-period $T$. The red line in Fig. \ref{fig_DOS}(a) shows the DOS under periodic boundary condition (PBC), i.e. $z=N_z+1$ and $z=1$ are taken as the same lattice site. Under such a condition, the DOS is solely contributed by the nodal loop states, hence it increases continuously with $T$ as more nodal loops with higher $n$ emerge when $T$ gets larger.
As a comparison, the blue line is the DOS obtained under OBC, which takes into account also the surface states. We can see that while the surface states have significantly increased the total density of states, the difference in DOS under PBC and OBC is almost unchanged. This is because the quantity of surface states of a NLSM is determined by the number of nodal loops and their winding directions. In our system, the loops with different $n$ take alternative winding directions (as shown by the winding number $\nu_1$ and $\nu_2$), thus they may either add together or cancel each other when determining the quantity of surface states. As a result, the surface DOS shall have some oscillating behavior with an increasing $T$. In Fig. \ref{fig_DOS}(b) we illustrate the SDOS as a function of $T$, which show a clear oscillating behavior.}

 {These changes of DOS can be reflected by some physical observables, e.g. the conductance in two-terminal transport measurements for cold-atom systems~\cite{transport_coldatoms1,transport_coldatoms2}. In such measurements, the system of interest is connected to two particle reservoirs on different sides, with a bias introduced between them. The bias takes the form of a chemical potential difference, which can be generated by different particle numbers of the two reservoirs. A ``wall" beam is introduced in addition to separate the two reservoirs. By turning off and on the ``wall" beam, one can control the length of time for the particles to move from one reservoir to another through the sandwiched system. After the ``wall" beam is turned back on, the particle number in the reservoirs can be measured using the time-of-flight technique, and the corresponding conductance can be extracted from a simple linear resistor-capacitor model \cite{transport_coldatoms2}.}
 {Note also that for periodically driven systems, the two-terminal conductance measurements must be slightly different  due to the lack of a Fermi surface in periodically driven systems. Fortunately, by now it is well established that surface state contributions to conductance at quasi-energy zero in periodically driven systems can be probed by transport measurements at energies at 0, $\pm \hbar\omega$, $\pm 2\hbar\omega$ ($\omega$ is the driving frequency) etc, the sum of their respective conductances will yield the conductance at quasi-energy zero \cite{transport_floquet1,transport_floquet2,transport_floquet3,Yap1,Yap2} and can hence tell the signatures of drumhead states. }    


\begin{thebibliography}{90}
\bibitem{footnote} The nodal loops wrapping around the 3-torus in the momentum space
 are exotic and interesting. But as far as the linkage in the 3D space $R^3$ is concerned, we do not find any exotic linkage associated with such loops.
\end{thebibliography}
\end{document}